\documentclass[twocolumn]{aastex631}

\usepackage{natbib}
\usepackage{hyperref}
\usepackage{appendix}
\usepackage{amsmath}

\usepackage[T1]{fontenc}

\begin{document}

\title{Little Red Dots on FIRE: The Ability of Bursty Galaxies to Host an Abundant Population of High-Redshift AGN}

\author[0000-0001-6676-4132]{Andrew Marszewski\footnotemark}
\affiliation{CIERA and Department of Physics and Astronomy, Northwestern University, Evanston, IL 60201}

\author[0000-0002-4900-6628]{Claude-André Faucher-Giguère}
\affiliation{CIERA and Department of Physics and Astronomy, Northwestern University, Evanston, IL 60201}

\author[0000-0003-4070-497X]{Guochao Sun}
\affiliation{CIERA and Department of Physics and Astronomy, Northwestern University, Evanston, IL 60201}

\author[0000-0001-5769-4945]{Daniel Anglés-Alcázar}
\affiliation{Department of Physics, University of Connecticut, 196 Auditorium Road, U-3046, Storrs, CT 06269-3046, USA}

\author[0000-0002-1109-1919]{Robert Feldmann}
\affiliation{Department of Astrophysics, University of Zurich, Zurich CH-8057, Switzerland}

\author[0000-0003-1598-0083]{Kung-Yi Su}
\affiliation{CIERA and Department of Physics and Astronomy, Northwestern University, Evanston, IL 60201}

\author[0000-0001-8367-6265]{Tim B. Miller}
\affiliation{CIERA and Department of Physics and Astronomy, Northwestern University, Evanston, IL 60201}

\author[0000-0002-0487-3090]{Niranjan Chandra Roy}
\affiliation{Department of Physics, University of Connecticut, 196 Auditorium Road, U-3046, Storrs, CT 06269-3046, USA}

\begin{abstract}
\footnotetext{Corresponding Author: Andrew Marszewski \\ \href{mailto:AndrewMarszewski2029@u.northwestern.edu}{AndrewMarszewski2029@u.northwestern.edu}}

The James Webb Space Telescope has unveiled an abundant population of potential active galactic nuclei (AGN) at high redshift ($z\gtrsim4$) known as little red dots (LRDs), which are likely hosted in relatively low-mass galaxies.  However, previous theoretical models have highlighted the difficulty in continuously feeding massive black holes in the central regions of bursty, high-redshift galaxies because of repeated gas evacuation by stellar feedback.  We analyze galaxies in high-redshift FIRE-2 simulations to understand whether they are capable of hosting the observed abundant population of high-redshift AGN. We use a gravitational torque-driven accretion (GTDA) model and a simple free-fall accretion model to derive black hole accretion rates and construct predicted AGN bolometric luminosity functions for $z=5-7$.  The GTDA model and the free-fall model with black holes accreting $\lesssim 1$ percent of their central gas supply ($<100 \rm \ pc$) per free-fall time predict AGN abundances that are more than sufficient to explain the most recent LRD observations.  The fiducial models, in fact, overpredict the number of low-luminosity AGN as compared with observations.  We explore possible resolutions of this tension.  A plausible, though likely not unique, scenario for alleviating the AGN overpredictions and which also provides a good match to the host-galaxy UV luminosity distribution suggests that LRDs are super Eddington-accreting, Eddington luminosity-limited, $M_{\rm BH}\gtrsim 2\times10^5 \ \rm M_\odot$ black holes residing in $M_\star\gtrsim 2\times10^7 \ \rm M_\odot$ galaxies.  We show that, under simple assumptions, mock observations of such sources can reproduce key observed LRD characteristics.

\end{abstract}

\section{Introduction} \label{sec:intro}

Observations with the James Webb Space Telescope (JWST) have revealed an abundant population of so-called little red dots (LRDs) in the early universe \citep{Matthee_2024}.  LRDs are characterized by their compactness and red rest-optical colors.  Some have argued that the spectra of LRDs may be stellar in origin (\citealp{Labbe_2023,Baggen_2024,Barro_2024}). This is in part due to their lack of classic AGN signatures (i.e., X-rays, e.g., \citealp{Yue_2024, Ananna_2024} and radio jets, e.g., \citealp{Akins_2025, Perger_2025}) and non-standard AGN SEDs.  However, there is increasing evidence that active galactic nuclei (AGN) contribute significantly to most, if not all, LRD spectra due to the high incidence of broad Balmer emission lines in LRD samples (\citealp{Harikane_2023, Kokorev_2023,Greene_2024, Furtak_2024, Matthee_2024}), observed time variability in some cases (\citealp{Furtak_2025}), and observations of Balmer breaks that are too strong to be explained by stellar populations alone (\citealp{Naidu_2025,deGraaff_2025a}).  Additionally, clustering constraints from \citet{Matthee_2025} suggest that LRD host galaxies have stellar masses that are too low ($M_\star \approx 5\times10^7 \ \rm M_\odot$) to reproduce their spectra with stellar emission alone.  Moreover, recent theoretical models plausibly explain the puzzling SEDs observed in LRDs in black hole accretion scenarios (e.g., \citealp{Madau_2025,Inayoshi_2025,Inayoshi_2025b}).

Under the interpretation that LRDs are AGN-dominated, standard bolometric luminosity calibrations find that their number density would exceed expectations derived from extrapolation of the quasar UV luminosity function by an order of magnitude (\citealp{Matthee_2024, Pizzati_2025}).  Furthermore, results from LRD censuses would imply greatly elevated AGN bolometric luminosity functions and black hole mass functions as compared to pre-JWST results (\citealp{Kokorev_2024, Kocevski_2025, Labbe_2025}).  Updated bolometric correction factors from \citet{Greene_2025}, driven largely by a lack of evidence for significant dust-reprocessed emission in the far-infrared (e.g., \citealp{Setton_2025,Casey_2025}), revise these estimates downward, potentially alleviating tension between theory and observed LRD luminosities and abundances.    

Previous simulation-based work has found that it is difficult to power AGN at high redshift due to stellar feedback regularly ejecting the central gas supply of low-mass galaxies (\citealp{Dubois_2015, Habouzit_2017, Habouzit_2021, Bower_2017, Lapiner_2021}).  At face value, this theoretical result is seemingly at odds with the observations of an abundant population of AGN at high redshift, a problem exacerbated by the discovery of numerous LRDs.

The FIRE (Feedback in Realistic Environments)\footnote{See the FIRE project website: \url{http://fire.northwestern.edu}.} project is a set of cosmological zoom-in simulations that resolve the multiphase ISM of galaxies and implement detailed models for star formation and stellar feedback (\citealp{Hopkins_2014,Hopkins_2018, Hopkins_2023}).  In agreement with other theoretical models, \citet{AA_2017b}, \citet{Catmabacak_2022}, and \citet{Byrne_2023} have previously shown that it is difficult to efficiently grow black holes in low-mass FIRE galaxies (which dominate at high redshift) due to bursty stellar feedback regularly evacuating gas from their centers.  The fiducial models in these works show that SMBH growth in FIRE simulations takes place in two phases when starting from stellar seeds: an early, highly intermittent phase where black holes are under-massive relative to local scaling relations and a later, more steady phase where black holes grow at higher time-averaged rates and converge to standard scaling relations.

With this motivation, we explore whether the central gas supplies of galaxies evolved using the FIRE model imply black hole accretion rates capable of driving the observed populations of LRDs and classical AGN at high redshift.  In particular, we aim to determine whether the intermittent feeding of black holes in high-resolution simulations of bursty, high-redshift, low-mass galaxies is in tension with the observed abundance of high-redshift AGN.  In a related study, \citet{Niranjan_2026} analyze massive galaxies in FIRE-2 cosmological simulations to investigate the extent to which a purely stellar ultra-compact phase can account for the observed properties of LRDs. They show that, although such galaxies can reproduce several key LRD-like features, such as their compactness, strong Balmer break, and broad Balmer lines due to galaxy-scale dynamics, additional physical processes, which are most likely AGN-driven, are needed to explain the full set of observed characteristics in the LRD population.  

This paper is organized as follows.  In Section \ref{sec:methods} we describe the high-redshift suite of FIRE-2 simulations analyzed in this work.  We present the different models used to measure accretion rates for hypothetical black holes placed at the centers of our simulated galaxies.  We also present our method for deriving a predicted AGN bolometric luminosity function from these measurements.  In Section \ref{sec:results} we present the AGN bolometric luminosity function predicted from our analysis for $z=5-7$ and compare with available results from high-redshift observational AGN censuses, including LRDs under the assumption that they are primarily powered by black hole accretion.  We study the dependence of our results on different model and parameter choices used to infer black hole accretion rates from our simulations.  In Section \ref{sec:discussion} we discuss our results and the feasibility of bursty, high-redshift FIRE-2 galaxies hosting the abundant population of LRDs unveiled by JWST.  We assess the extent to which different modifications to our fiducial models can improve agreement between our predictions and observed LRD demographics.  Our results support the possibility that LRDs host AGN accreting at super-Eddington rates.  Finally, we summarize the main conclusions of this work in Section \ref{sec:conclusions}.  In Appendices \ref{appendix:radial_dependence}, \ref{appendix:MBH_Assumption}, and \ref{appendix:Dust} we explore the dependence of our results on the radial aperture used in our accretion rate calculations, our assumed black hole to stellar mass relation, and our assumption of dust attenuation, respectively.

\begin{figure*}[!t]
    \centering
    \includegraphics[width=\linewidth]{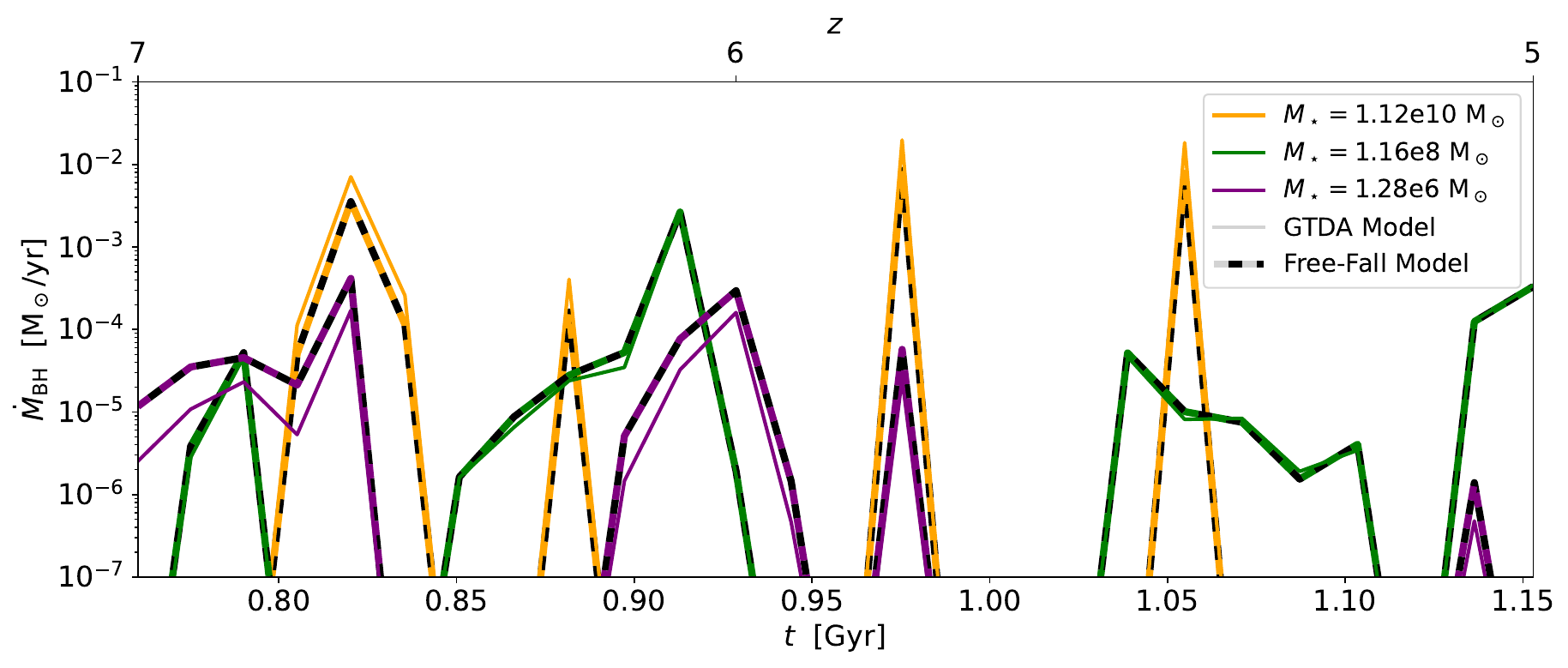}
    \caption{Predicted black hole accretion rates using the gravitational torque-driven accretion (GTDA) model (solid lines) and the free-fall accretion model with $\varepsilon_{\rm ff}=0.001$ (dashed lines) for three example host galaxies in our sample with $M_{\star}(z=5) \sim 10^6$ (purple), $10^8$ (green), and $10^{10} \ (\rm orange) \ \rm M_\odot$ for $z=5-7$.  The horizontal axis shows cosmic time.  Bursty stellar feedback in FIRE-2 (and in many other high-resolution simulations) disrupts the central gas supply of galaxies and prevents black holes from continuously accreting at high redshift.  This effect potentially drives tension between theoretical models and the observed abundance of AGN at high redshift.}
    \label{fig:Accretion_Time_Series}
\end{figure*}

\section{Methods} \label{sec:methods}

\subsection{The Simulations}

We analyze a high-redshift suite of FIRE-2 cosmological zoom-in simulations originally presented by \citet{Ma_2018a,Ma_2018b,Ma_2019}. These have been extensively validated against high-redshift observations and provide a robust foundation for this work.  Notably, \citet{Sun_2023b} demonstrate that this simulation suite reproduces the $z=8-12$ UV luminosity function (UVLF) measured by JWST.  Furthermore, the same FIRE-2 physics model also underlies the FIREbox$^{\it HR}$ cosmological volume simulation which provides a good match to both the observed UVLF and UV luminosity density over $z\sim{}6-14$ \citep{Feldmann_2025}.  \citet{Marszewski_2024, Marszewski_2025} use the high-redshift, zoom-in suite of FIRE-2 simulations to characterize and explain the form and weak evolution of the gas-phase mass-metallicity relation for $z \geq 5$ and show that it is in excellent agreement with JWST measurements.  

This simulation suite was run using the GIZMO code \citep{Hopkins_2015}.  The hydrodynamic equations are solved using GIZMO's meshless finite-mass (MFM) method.  The 20 particular simulations analyzed in this paper are the z5m12a--e, z5m11a--i, and z5m10a--f runs.  The names of these simulations denote the final redshift that they were run down to ($z_{\rm fin}=5$) and the main halo masses (ranging from $M_{\rm halo} \approx 10^{10}-10^{12} \ \rm{M}_\odot$) at this final redshift.  Baryonic (gas and star) particles have initial masses $m_b = 900-7000 \ \rm{M}_\odot$ (simulations with more massive main halos have more massive baryonic particles).  Dark matter particles are more massive by a factor of $\Omega_{\rm DM}/\Omega_{\rm b} \approx 5$. Gravitational softenings are adaptive for the gas and are fixed to $\epsilon_* = 1.4-2.1$ physical pc and $\epsilon_{\rm DM} = 21-42$ physical pc for star and dark matter particles, respectively.

Notably, these simulations do not include black hole physics and AGN feedback.  In the next section we describe the different methods we use to calculate inferred accretion rates onto hypothetical black holes placed at the center of our galaxies.  As shown in Figure \ref{fig:Accretion_Time_Series}, inferred black hole accretion rates in these galaxies are highly intermittent due to the bursty stellar feedback, which is a prediction of the FIRE-2 model for galaxies in this mass ($M_{\star} \lesssim 10^{10} \ \rm{M}_\odot$) and redshift range, regularly evacuating their central gas supply.  We will discuss in Section \ref{sec:limitations} the potential effects of AGN feedback neglected here.  

\subsection{Predicting Accretion Rates} \label{sec:Measure_Accretion}

We use both the gravitational torque-driven accretion (GTDA) model and a simple free-fall accretion model to predict instantaneous accretion rates onto hypothetical black holes placed at the centers of our galaxies.  In both cases, we use Amiga Halo Finder (AHF; \citealp{Knollmann_2009}) to identify halos in our zoom-in simulations.  We place black holes at the point of maximum stellar density within one $R_{\rm vir}$ of the center of mass of each halo identified by AHF.  We then measure physical quantities within a spherical aperture around the black hole and use these measurements to infer black hole accretion rates.  \citet{AA_2017b} previously compared black hole accretion rates calculated using this post-processing approach to those calculated on-the-fly and validated the post-processing approach.  Below we describe the two models used in this work to calculate accretion rates in post-processing.

\begin{figure*}[!t]
    \centering
    \includegraphics[width=\linewidth]{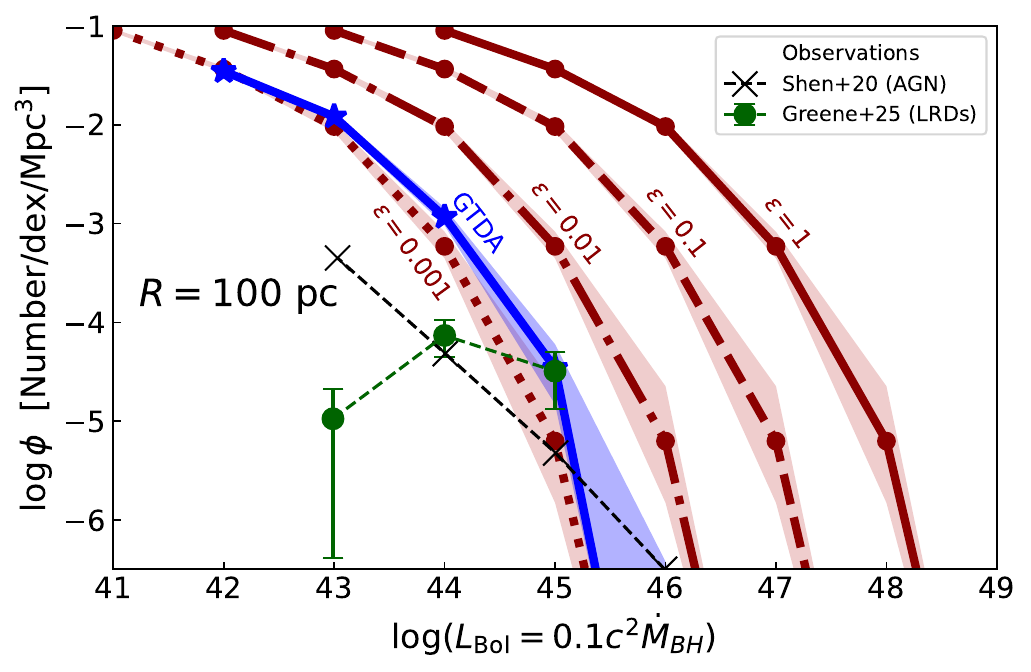}
    \caption{The AGN bolometric luminosity function predicted at $z \sim 6$ using the GTDA model (blue with stars) and the simple free-fall accretion model (dark red with circles) with $\bar\varepsilon_{\rm ff} = 1$ (solid), $0.1$ (dashed), $0.01$ (dot-dashed), and $0.001$ (dotted).  Curves show the median luminosity function predicted from our 1000 bootstrapped samples; shaded regions enclose the 16th–84th percentile range.  Here, our fiducial aperture of $R=100 \ \rm pc$ is used and we introduce our fiducial amount of variance to the normalization/accretion efficiency ($\sigma_{\log\epsilon_T} = \sigma_{\log\varepsilon_{\rm ff}}=0.5$).  The inferred accretion rates (and therefore the predicted AGN bolometric luminosity function) for the GTDA model and the simple free-fall accretion model with $\bar\varepsilon_{\rm ff}\sim0.001$ are in close agreement.  Green circles with error bars show the LRD bolometric luminosity function inferred from observations by \citet{Greene_2025} for $z=4-6$ and black crosses show the pre-JWST luminosity function based on X-ray and UV observations from \citet{Shen_2020} (their ``Global Fit A'') for $z=4-6$.  Our simple free-fall accretion model with an accretion efficiency of less than one percent is able to reproduce the abundance of objects in the highest observed luminosity bin ($L_{\rm Bol} \sim 10^{45} \ \rm erg/s$) presented by \citet{Greene_2025} but overpredicts the number of AGN in lower-luminosity bins.}
    \label{fig:AGN_Bol_LF_vs_epsilon}
\end{figure*}

\subsubsection{Gravitational Torque-Driven Accretion}

We estimate black hole accretion rates using the GTDA model introduced by \citet{Hopkins_2011}.  This model is the default accretion model used in FIRE simulations that include live black holes (e.g., \citealp{AA_2017b, Wellons_2023,Byrne_2024}).  In this model, the accretion rate is given by 

\begin{equation} \label{eqn:GTDA}
    \begin{aligned}
    \dot{M}_{\rm Torque} = & \epsilon_{\rm T}f_{\rm d}(<R)^{5/2}\left(\frac{M_{\rm BH}}{10^8 \rm M_\odot}\right)^{1/6} \\ &\times\left(\frac{M_{\rm tot, b}(<R)}{10^9 \rm M_\odot}\right) \left(\frac{R}{100 \ \rm pc}\right)^{-3/2} \\ &\times \left(1+\frac{f_0(<R)}{f_{\rm gas}(<R)}\right)^{-1} \ \frac{\rm M_\odot}{\rm yr},
    \end{aligned}
\end{equation}
where
\begin{equation} 
    f_0 \approx 0.31f_d(<R)^2\left(\frac{M_{\rm d}(<R)}{10^9\rm M_\odot}\right)^{-1/3},
\end{equation}
and $\epsilon_T$ is a constant factor that controls the normalization of the BH-galaxy scaling relations in torque-limited accretion scenarios (\citealp{AA_2015,AA_2017c}).  Throughout this work we choose a fiducial mean value for $\bar\epsilon_T$ of 5 (on the high side of values shown by \citet{AA_2017c} to reproduce a reasonable $z=0$ $M_{\rm BH} - M_\star$ scaling relation), but we explore the effects of introducing log-normal scatter in this parameter.  Here and throughout most of this paper, we choose our black hole masses to be $M_{\rm BH} = 0.01M_\star$ (one percent of the galaxy’s stellar mass within $0.2R_{\rm vir}$ of its center), corresponding to moderately overmassive black holes relative to local scaling relations and in approximate agreement with the mean scaling relation presented by \citet{Harikane_2023} for $z=4-7$ galaxies.  Note the weak scaling of $\dot{M}_{\rm Torque}$ with $M_{\rm BH}$ makes accretion rates predicted from the GTDA model relatively insensitive to this choice in black hole mass.  The argument $(<R)$ indicates that a quantity is calculated using particles within a radius $R$ from the galaxy's center.  We use the fiducial value of $R = 100 \ \rm pc$ for reasons discussed in Appendix \ref{appendix:radial_dependence} where we show the dependence of the predicted bolometric AGN luminosity function on $R$.  Here, $M_{\rm tot,b}$ is the total baryonic mass (stars and gas), $M_{\rm d}$ is the mass of stars and gas in the disk, and $f_{\rm d} = M_{\rm d}/M_{\rm tot,b}$ is the disk fraction.  The disk mass is calculated as $M_{\rm d} = M_{\rm tot, b} - M_{\rm bulge}$, where $M_{\rm bulge}$ is estimated as twice the baryonic mass that is counter-rotating relative to the net angular momentum of all baryonic mass within $R$ from the galaxy's center.  In Appendix \ref{appendix:MBH_Assumption}, we explore the effects of varying our assumed $M_{\rm BH} - M\star$ relation and the self-consistency of our fiducial relation. 

\subsubsection{Simple Free-Fall Accretion}
We also use a simple free-fall accretion model to predict black hole accretion rates for our galaxies.  In this model the accretion rate is given by
\begin{equation} \label{eqn:simple_accretion}
    \dot{M}_{\rm BH} = \varepsilon_{\rm ff}\frac{M_{\rm gas}(<R)}{t_{\rm ff}},
\end{equation}
where $M_{\rm gas}(<R)$ is the gas mass located within an aperture of radius $R$ from the galaxy's center and 
\begin{equation}
    t_{\rm ff} \equiv \sqrt{\frac{\pi^2R^3}{ 8GM_{\rm tot}(<R)}}
\end{equation}
is the free-fall time of that gas.  Here, $M_{\rm tot}(<R)$ is the total (gas, stellar, black hole, and dark matter) mass within an aperture of radius $R$ from the galaxy's center.  Note that our predicted accretion rates with this model are again relatively insensitive to our choice in black hole mass because $M_{\rm BH}$ is usually a subdominant contribution to $M_{\rm tot}$, with a maximum scaling of $\dot{M}_{\rm BH} \propto M_{\rm BH}^{1/2}$ in the extreme scenario that a black hole dominates the total mass of its galaxy within $R = 100 \ \rm pc$.  Finally, $\varepsilon_{\rm ff}$ is the dimensionless efficiency for which gas within this aperture accretes onto the black hole per free-fall time.  Values for $\varepsilon_{\rm ff}$ can, in principle, range between 0 and 1.  We choose a fiducial mean value of $\bar\varepsilon_{\rm ff}=0.01$. We explore the effects of introducing log-normal scatter to this quantity while keeping the mean efficiency fixed. 

The major benefits of this model are its simplicity and interpretability since the parameter $\varepsilon_{\rm ff}$ tells us what percentage of gas within our radial aperture is accreted by the black hole per free-fall time.  A major motivation for this work is that, a priori, it was not clear that any value of $\varepsilon_{\rm ff} \leq 1$ would predict accretion rates sufficient to reproduce LRD observations.  However, as we show in Section \ref{sec:results}, a mean efficiency of $\bar\varepsilon_{\rm ff} \lesssim 0.01$ is sufficient to reproduce observations in the highest observed LRD luminosity bins. 

\begin{figure*}[!t]
    \centering
    \includegraphics[width=\linewidth]{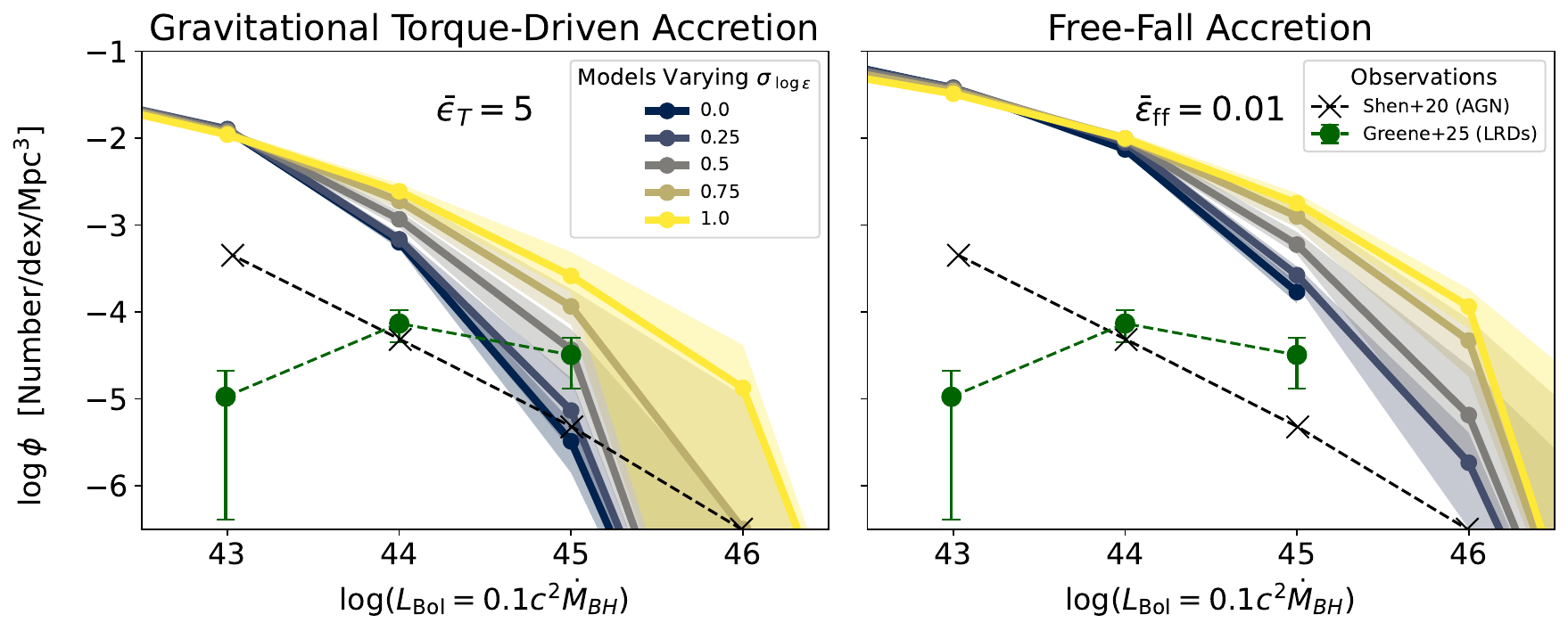}
    \caption{The predicted $z \sim 6$ AGN bolometric luminosity function from FIRE-2 simulations using the GTDA model (left panel) and the simple free-fall accretion model (right panel).  To model unresolved time variability, we apply log-normally distributed values of the normalization ($\bar\epsilon_T=5$) for the GTDA model and the accretion efficiency ($\bar\varepsilon_{\rm ff}$=0.01) for the simple free-fall model with $\sigma_{\log\epsilon} = 0$ (dark blue), $0.25$ (blue-gray), $0.5$ (gray), $0.75$ (gold), and $1$ (yellow).  Shaded regions represent the range between the 16th and 84th percentile luminosity functions predicted from our bootstrapped samples.  Here, our fiducial aperture of $R = 100 \ \rm pc$ is used. Green circles with error bars show the LRD bolometric luminosity function inferred from observations by \citet{Greene_2025} for $z=4-6$ which goes through the pre-JWST luminosity function based on X-ray and UV observations from \citet{Shen_2020} (their ``Global Fit A'') represented by black crosses.  Introducing scatter to the normalizations/accretion efficiencies upscatters some low-luminosity sources into higher-luminosity bins, thereby making the slope of the predicted luminosity function shallower and in better agreement with observations in the higher luminosity bins.  In all cases, however, our models overpredict the number of faint AGN as compared with observations.}
    \label{fig:AGN_Bol_LF_vs_sigma}
\end{figure*}

\subsection{Introducing Subgrid Variability in Accretion Rates} \label{sec:Introducing_Variation}

The models described above provide predictions for accretion rates onto the black holes after averaging galaxy properties over apertures of radius $R=100 \ \rm pc$ (in the fiducial analysis).  Due to variations in the gas dynamics from galaxy to galaxy and the time-variable nature of black hole accretion for individual galaxies, we expect there is some unresolved variation in these predicted accretion rates.  Indeed, \citet{AA_2021} and \citet{Hopkins_2024a} apply hyper-refinement to cosmological simulations to study black hole accretion flows on smaller scales and find that accretion rates vary significantly on timescales shorter than can be resolved by applying models to physical properties within a $100 \ \rm pc$ aperture.  We can model this variation as manifesting through variable values of the normalization factor $\epsilon_T$ in the GTDA model or in the accretion efficiency $\varepsilon_{\rm ff}$ in the simple free-fall model.  In order to be relevant, the subgrid variability introduced here need only occur on timescales shorter than our snapshot resolution ($\Delta t_{\rm snapshot}\sim 10 \ \rm Myr$) which is much longer than the timescales for observations of LRD variability (typically days to years).  Thus, the inclusion of significant subgrid variability is not in tension with the relatively low amount of variability that has been observed in LRDs (e.g., \citealp{Zhang_2025obs,Tee_2025}).  Here, we describe our method for applying log-normally distributed values of $\epsilon_T$ and $\varepsilon_{\rm ff}$ to our sample.

For the GTDA model, we generate and apply (base-10) log-normal distributions of $\epsilon_T$ with mean values of $\bar\epsilon_T=5$ and standard deviations of $\sigma_{\log (\epsilon_{T})} = 0, \ 0.25,\ 0.5, \ 0.75, \ \rm and \ 1$. For the simple free-fall accretion model, we apply (base-10) log-normal distributions with mean accretion efficiencies of $\bar\varepsilon_{\rm ff}=0.01$ and standard deviations of $\sigma_{\log (\varepsilon_{\rm ff})} = 0, \ 0.25,\ 0.5, \ 0.75, \ \rm and \ 1$.  Note that we preserve the linear average efficiency when injecting different levels of log scatter by offsetting the center of the log-normal distribution by $-\sigma^2/2$.  

From visual inspection of results from the hyper-refined simulations in \citet{AA_2021} and \citet{Hopkins_2024a}, we find that accretion rates typically vary by about half an order of magnitude ($\sigma_{\log (\epsilon)}\sim0.5$) on timescales shorter than are resolved here.  This value is based on very limited examples.  We therefore select our fiducial amount of injected variation in accretion rates to be $\sigma_{\log (\epsilon_{T})}=\sigma_{\log (\epsilon_{\rm ff})}=0.5$ but vary this parameter to explore its effects on our results.  The $\sigma_{\log (\epsilon)}=0$ case corresponds to constant normalizations/efficiencies while larger values of $\sigma_{\log (\epsilon)}$ correspond to larger amounts of assumed, unresolved variation in accretion rates.  Note that these randomly generated distributions are regenerated and reapplied to our accretion rates for each resampling performed using the bootstrapping method described in the final paragraph of section \ref{sec:Calc_L_func}.  Applying a newly generated distribution for each bootstrapped resample mitigates the level of random noise introduced by a single realization of the log-normal distributions.  Figure \ref{fig:AGN_Bol_LF_vs_sigma} shows the effect of applying log-normally distributed normalizations/efficiencies with different variances on the predicted AGN bolometric luminosity function.

\subsection{The AGN Bolometric Luminosity Function} \label{sec:Calc_L_func}

We use our inferred accretion rates to predict the AGN bolometric luminosity function.  To begin, we calculate the expected bolometric luminosity for the black hole at the center of each galaxy in our sample using
\begin{equation} 
    L_{\rm Bol} = \eta \dot{M}_{\rm BH}c^2.
\end{equation}
Throughout this work, we assume a standard black hole radiative efficiency of $\eta=0.1$, expected for radiatively efficiently accreting, moderately spinning ($a\sim0.5-0.7$) supermassive black holes (\citealp{Soltan_1982,Yu_2002}). Note this could be up to $\eta \sim 0.4$ for maximally spinning black holes (\citealp{NovikovThorne1973,PageThorne1974}).  

The volumes captured by our zoom-in simulations are not representative volumes of the universe and therefore cannot be used to directly predict the luminosity function. To address this, we convolve our resulting AGN bolometric luminosities with the halo mass function (HMF) using the ``HMF-weighting'' method originally used by \citet{Ma_2018b} to analyze the UVLF for a subset of the simulations in the present paper.  \citet{Sun_2023b} also use this method to calculate the UVLF for the full suite of simulations used here and show that it is consistent with JWST observations of the UVLF for $z=8-12$.  This method works by first dividing halos into narrow mass bins.  Halos within the same mass bin are all given equal weight $w=N_{\rm E} / N_{\rm S}$, where $N_{\rm S}$ is the number of simulated halos in that mass bin and $N_{\rm E}$ is the number of halos in that mass bin expected by the HMF, calculated using the HMF from \citet{Behroozi_2013} (appropriate at the high redshifts of interest for this work) with code from \citet{Murray_2013}. These weights are then applied to halos as they are binned by $L_{\rm Bol}$ to construct the AGN bolometric luminosity function.  See Section 2.4 in \citet{Ma_2018b} for a full description of this method.

We use the bootstrapping method to estimate the error bars of our predicted AGN bolometric luminosity functions.  That is, we perform random resampling (with replacement) of the accretion rates used in our calculation of the luminosity function 1000 times.  Each time, the sample drawn is the same size as our original sample and a predicted bolometric luminosity function is calculated from that sample.  In all plots of the luminosity function, we present the median luminosity function calculated from these 1000 realizations and shaded regions (when plotted) represent the range between the 16th and 84th percentiles of these luminosity functions.  New log-normally distributed values of $\epsilon_T$ and $\varepsilon_{\rm ff}$ (discussed in section \ref{sec:Introducing_Variation}) are generated and applied with each resampling, minimizing the random noise introduced by any one randomly generated distribution.

\section{Results} \label{sec:results}

Here we present the AGN bolometric luminosity functions predicted from our black hole accretion models and test the sensitivity of our results to various parameters in these models.  We begin by calculating the bolometric luminosity function using accretion rates derived from the GTDA and simple free-fall accretion models based on the properties of particles within a fiducial aperture of $R=100 \rm \ pc$.  Appendix \ref{appendix:radial_dependence} investigates the dependence of our results on this choice in radial aperture.  

\subsection{Bolometric Luminosity Function More than Explained with Modest Accretion Efficiencies} \label{sec:Epsilon}

For the GTDA model, we fix our mean normalization at $\bar\epsilon_{\rm T}=5$.  Using the free-fall accretion model, we apply different mean values of the accretion efficiency, $\bar\varepsilon_{\rm ff}$, to understand the efficiency at which black holes would have to accrete their central gas supply in order to reproduce the observed AGN bolometric luminosity function.  In both cases, we apply our fiducial amount of introduced log-normal variance, $\sigma_{\log (\epsilon_{\rm T})}=\sigma_{\log (\varepsilon_{\rm ff})}=0.5$.  Figure \ref{fig:AGN_Bol_LF_vs_epsilon} compares the AGN bolometric luminosity functions predicted from the GTDA model and the simple free-fall accretion model using $\bar\varepsilon_{\rm ff} = 0.001,\ 0.01,\ 0.1,\ \rm and \ 1$ with observations.

We find that black holes typically accreting $\lesssim 1$ percent of their gas supply within $R=100 \rm \ pc$ per free-fall time is more than sufficient to reproduce the abundance of sources in the most luminous bin associated with LRDs ($L_{\rm Bol} \sim 10^{45} \ \rm erg/s$) observed by \citet{Greene_2025}.  Note that updated bolometric correction factors from \citet{Greene_2025} significantly revised the LRD bolometric luminosity function toward fainter luminosities relative to previous JWST estimates (e.g., \citealp{Kokorev_2024,Kocevski_2025,Labbe_2025}), making it easier to explain with theory and more closely matching the pre-JWST AGN luminosity function from \citet{Shen_2020} (their ``Global Fit A'').  The comparison with \citet{Shen_2020} is presented differently in \citet{Greene_2025} since they compare with ``Global Fit B'' from \citet{Shen_2020}, which is derived from an X-ray luminosity function that is now known to be incomplete at the faint end due to the faintness of LRDs in the X-ray (e.g., \citealp{Yue_2024,Mailano_2025}). 

Interestingly, our fiducial GTDA model and free-fall accretion model with $\bar\varepsilon_{\rm ff} = 0.01$ overpredict the amount of sources observed at lower luminosities ($L_{\rm Bol} \sim 10^{43-44} \ \rm erg/s$).  In section \ref{sec:Match_Obs} we discuss potential explanations for this discrepancy.

\subsection{Effects of Subgrid Variability} \label{sec:variable_eps}
  
Here, we investigate the effects of applying different log-normal distributions of $\epsilon_T$ and $\varepsilon_{\rm ff}$ to the accretion rates inferred from the GTDA and free-fall accretion models, respectively, on the predicted bolometric luminosity function.  For the GTDA model, we apply (base-10) log-normal distributions with mean normalizations of $\bar\epsilon_{\rm T}=5$ and standard deviations of $\sigma_{\log (\epsilon_{\rm T})} = 0, \ 0.25,\ 0.5, \ 0.75, \ \rm and \ 1$.  Likewise, for the simple free-fall accretion model, we apply (base-10) log-normal distributions with mean accretion efficiencies of $\bar\varepsilon_{\rm ff}=0.01$ and standard deviations of $\sigma_{\log (\varepsilon_{\rm ff})} = 0, \ 0.25,\ 0.5, \ 0.75, \ \rm and \ 1$.  Figure \ref{fig:AGN_Bol_LF_vs_sigma} shows the effect of applying log-normally distributed normalizations or efficiencies with different variances on the predicted AGN bolometric luminosity function.

The application of log-normally distributed efficiencies/normalizations upscatters some low-luminosity sources into higher-luminosity bins.  While some high-luminosity sources are also downscattered into lower-luminosity bins, the larger abundance of lower-luminosity sources generates a net upscattering effect.  This asymmetric scattering effect allows our models to more easily reproduce the abundance of the most luminous observed sources.  This also results in shallower slopes of the predicted bolometric luminosity function in better agreement with observations, although, the predicted abundance of low-luminosity AGN still far exceeds what is present in observations.  Potential explanations for this discrepancy are discussed in Section \ref{sec:Match_Obs}.

\section{Discussion} \label{sec:discussion} 

\subsection{Bursty Galaxies Predicted to Host Ample AGN Population} \label{sec:alleviated_tension}

We have demonstrated that accretion rates derived from applying the GTDA model and a simple free-fall accretion model to the central regions of FIRE-2 galaxies imply AGN abundances that are more than sufficient to explain high-redshift observations.  This demonstrates that the abundant population of likely AGN observed at high redshift in the form of LRDs is not in tension with the intermittent black hole feeding at high redshift in FIRE simulations.  Despite regular disruption of the central gas supply by bursty stellar feedback, FIRE-2 galaxies are predicted to host accreting black holes frequently enough such that their population sustains a substantial abundance of AGN.  This result is reassuring given that, in addition to the FIRE simulations, many other high-resolution simulations of high-redshift galaxies predict intermittent black hole feeding. 

As shown in Figure \ref{fig:AGN_Bol_LF_vs_epsilon}, our fiducial GTDA model can reproduce the number of AGN observed in the highest luminosity bin presented by \citet{Greene_2025} and actually overpredicts the number density observed in lower luminosity bins.  Our fiducial simple free-fall accretion model ($\bar\varepsilon_{\rm ff}=0.01$) overpredicts the number of AGN in all luminosity bins presented by \citet{Greene_2025}.  This indicates that black holes accreting an average of one percent of their central ($<100 \rm \ pc$) gas supply per free-fall time is more than sufficient to reproduce LRD observations.  Note that this overprediction is not unique to our work and is seen in other theoretical models, including models from \citet{Volonteri_2017} and the DELPHI models \citep{Dayal_2025}.  We now explore how different cuts imposed on our AGN and their host galaxies could allow our theoretical predictions to more closely match the observed population of LRDs.

\begin{figure*}[!t]
    \centering
    \includegraphics[width=\linewidth]{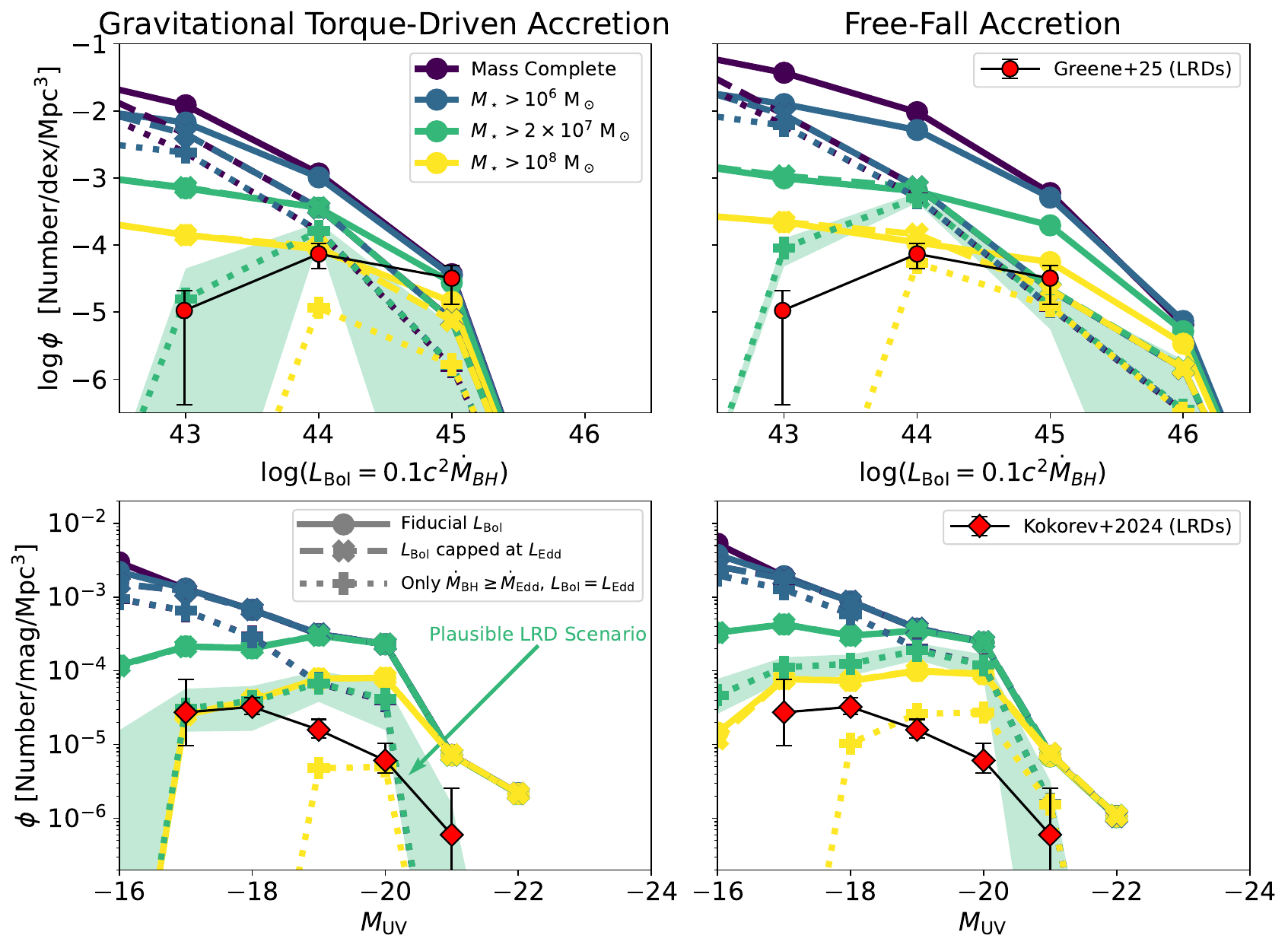}
    \caption{The predicted $z \sim 6$ AGN bolometric luminosity function (top) and UVLF for galaxies hosting an AGN in the LRD luminosity bins ($L_{\rm Bol} \sim 10^{43-45} \ \rm erg/s$; bottom) using the GTDA model (left) and the simple free-fall accretion model (right).  Galaxies only contribute to the UVLFs plotted in the bottom panels if they are predicted (by the model plotted) to host an AGN in the range $L_{\rm Bol} = 10^{42.5-45.5} \ \rm erg/s$.  Since not all AGN are LRDs, we make different modifications to our fiducial model in an attempt to better match observations of the LRD bolometric luminosity function from \citet{Greene_2025} (red circles) and the $z=4.5-6.5$ LRD UVLF from \citet{Kokorev_2024} (red diamonds).  In these models, the UV luminosity is powered by star formation in the host galaxy only.  We present the cases where galaxies of any mass are permitted to host AGN (dark purple) and where the contributions of AGN hosted within galaxies with $M_\star<10^{6} \ \rm M_{\odot}$ (blue), $M_\star<2\times10^{7} \ \rm M_{\odot}$ (green), and $M_\star<10^{8} \ \rm M_{\odot}$ (yellow) are excluded.  With these host galaxy stellar mass cuts, we simultaneously explore the cases where the luminosities of AGN are capped at their predicted $L_{\rm Edd}$ (dashed) and where we only include super-Eddington accretors whose luminosities are still capped at $L_{\rm Edd}$ (dotted).  The model where we include only super-Eddington-accreting, Eddington luminosity-limited black holes hosted within $M_\star>2\times10^{7} \ \rm M_{\odot}$ galaxies (our ``Plausible LRD Scenario''; dotted green line with shaded region representing the range between the 16th and 84th percentile luminosity functions from our bootstrapped samples) is an example of a plausible scenario that comes close to simultaneously reproducing the observed LRD bolometric and LRD UV luminosity functions.}
    \label{fig:LowL_Tension}
\end{figure*}

\subsection{Matching LRD AGN and Host Galaxy Demographics} \label{sec:Match_Obs}

We discuss and investigate potential reasons for our fiducial models' overprediction of the number of AGN in the lower luminosity bins associated with LRDs ($L_{\rm Bol} \sim 10^{43-44} \rm \ erg/s$) as compared with observations.  One certain difference between our predicted AGN bolometric luminosity function and that observed by \citet{Greene_2025} is that our model predicts the abundance of all AGN at a given luminosity while their sample is made up of only objects selected as LRDs.  However, as shown in, e.g., \citet{Harikane_2023}, \citet{Hviding_2025}, and demonstrated by the similar magnitudes of the AGN luminosity function from \citet{Shen_2020} and LRD luminosity function from \citet{Greene_2025}, LRDs make up a significant fraction of all observed broad-line sources at $z\gtrsim4$.  The maximum AGN bolometric luminosity function including all broad-line sources would be only a factor of $\sim 2-3$ higher than that presented in \citet{Greene_2025} which would do little to eliminate the tension with our fiducial models at the low-luminosity end.

In addition to the LRD bolometric luminosity function, the LRD UVLF serves as an important constraint on LRD demographics, particularly on their host galaxies.  Note that recent observations (e.g., \citealp{Killi_2024, Naidu_2025}) and theoretical models (e.g., \citealp{Inayoshi_2025b}) suggest that UV emission from LRDs is dominated by star formation in their host galaxy.  The bottom panels of Figure \ref{fig:LowL_Tension} show the UVLF for galaxies that host AGN in the bolometric luminosity range of LRDs ($L_{\rm Bol} \sim 10^{43-45} \rm \ erg/s$) calculated using our fiducial models and modified models (see Section \ref{sec:mods}).  As a result of overpredicting the number of AGN in the luminosity range of LRDs, our fiducial models also drastically overpredict the $z=4.5-6.5$ UVLF of LRD host galaxies observed by \citet{Kokorev_2024}.  These fiducial models include all galaxies as potential AGN hosts and impose no Eddington limit on the AGN luminosities inferred from our accretion models.  

Below, we explore the ability of different sets of physically motivated modifications to our models to better match LRD demographics.  These modifications consist of hypotheses regarding which galaxies in our sample are capable of hosting AGN, the limits of their AGN luminosities, and the subset of these AGN that are LRDs.

\subsubsection{Motivated Modifications to Our Models} \label{sec:mods}

Here, we list, describe, and motivate the different sets of modifications we impose on the galaxies and predicted AGN included in our sample to better match LRD observations.  Throughout this discussion we define a galaxy's stellar mass $M_\star$ to be the stellar mass enclosed within $0.2R_{\rm vir}$ of its center and assume its black hole mass to be one percent of its stellar mass ($M_{\rm BH} = 0.01M_\star$).  In Appendix \ref{appendix:MBH_Assumption}, we explore the dependence of results from our ``Eddington Luminosity Cap'' and ``Super-Eddington Accretors'' scenarios (described below) on and the self-consistency of this assumed $M_{\rm BH} - M_\star$ relation.

\begin{itemize}
  \item \textbf{Minimum Black Hole/Host Galaxy Mass:}  We explore three different cuts where we exclude all contributions to our luminosity functions from host galaxies with $M_\star < 10^6 \ \rm M_\odot$, $M_\star <  2\times10^7 \ \rm M_\odot$, or $M_\star < 10^8 \ \rm M_\odot$ ($M_{\rm BH} < 10^4 \ \rm M_\odot$, $M_{\rm BH} <  2\times10^5 \ \rm M_\odot$, or $M_{\rm BH} < 10^6 \ \rm M_\odot$, respectively).  The exclusion of low-mass black holes/galaxies is potentially motivated by theoretical work (e.g., \citealt{Ma_2021}) that has shown low-mass black holes, likely to be hosted in low-mass galaxies, to have difficulty in migrating to the center of their host galaxy via dynamical friction (both because of the long dynamical friction timescales and because low-mass, high-redshift galaxies are often irregular, with no stable dynamical center).  If this is the case, then the true luminosities of low-mass black holes in low-mass galaxies would likely be much lower than predicted by our fiducial models which place black holes at the maximum stellar density of all galaxies.  Note that the most recent observational estimates from \citet{Greene_2025} predict that typical LRDs are powered by $M_{\rm BH} \sim 10^{5-7} \ \rm M_\odot$ black holes hosted within $M_\star \sim 10^{8} \ \rm M_\odot$ galaxies.  Moreover, clustering constraints presented by \citet{Matthee_2025} suggest stellar masses of $M_\star \approx 5\times10^{7} \ \rm M_\odot$ for broad-line H$\rm \alpha$ emitters (including LRDs) at z = 4–5.  However, LRDs may have a wide range of host galaxy stellar masses and stellar mass measurements are difficult under the interpretation that AGN emission dominates at wavelengths longer than the Balmer limit.
  
  \item \textbf{Eddington Luminosity Cap:} We test whether capping the bolometric AGN luminosities in our sample at the Eddington luminosity  ($L_{\rm Bol} \leq L_{\rm Edd} = 1.26 \times10^{42} \ (M_{\rm BH}/10^6 \ \rm M_\odot) \rm \ erg/s$) can alleviate the overprediction of our fiducial models.  This modification to our model is motivated by the fact that, while some black holes may accrete at super-Eddington rates, their luminosities are likely still limited near the Eddington luminosity or a factor of order unity above it (see the accretion flow simulations in e.g., \citealp{Ohsuga_2002, Jiang_2014, Sadowski_2014, Madau_2014, Zhang_2025}).  Note that applying this Eddington luminosity cap is equivalent to lowering the radiative efficiency, $\eta$, of super-Eddington accretors below the fiducial value of $\eta=0.1$.
  
  \item \textbf{Super-Eddington Accretors:} Finally, we test the effects of including only black holes that are inferred to accrete at super-Eddington rates ($\dot{M}_{\rm BH} > \dot{M}_{\rm Edd}$), but still have Eddington-limited luminosities ($L_{\rm Bol} = L_{\rm Edd}$).  Note that general relativistic magneto-hydrodynamic (GRMHD) simulations with radiation transport predict that black holes can achieve super-Eddington accretion rates (\citealp{Jiang_2014, Sadowski_2014}).  The motivations for requiring super-Eddington accretion in LRDs are twofold.  First, LRDs may need luminosities at or near $L_{\rm Edd}$ in order to explain the presence of such luminous AGN in low-mass galaxies (predicted to host low-mass black holes).  However, since the gravitational radii of influence of black holes are generally small compared to the size of their host galaxies, the rate at which galaxies funnel gas toward their central black hole is not tuned to the Eddington limit, which depends only on black hole mass.  As a result, we expect black holes with luminosities near $L_{\rm Edd}$ to be supplied with gas at super-Eddington rates frequently.  Secondly, recent theoretical models (e.g., \citealp{Madau_2024, Madau_2025, Inayoshi_2025, Inayoshi_2025c}) suggest that super-Eddington accretion may also be needed in order to explain the puzzling shapes of LRD SEDs.  Section 4.7 in \citet{Zhang_2025} use GRMHD simulations to illustrate how super-Eddington accretion can result in both the weak X-ray emission and strong Balmer breaks observed in LRDs.
\end{itemize}

Note that there are many uncertainties introduced by assumptions in our modeling (e.g., the aforementioned assumed $M_{\rm BH} - M_\star$ relation; see Appendix \ref{appendix:MBH_Assumption}), different modifications one could explore, and potential degenerate effects of these modifications.  Therefore, the modifications described here are only meant to illustrate the ability of plausible modifications to bring our results into better agreement with observations rather than to provide hard constraints on the nature of LRDs.

\subsubsection{Predicting the LRD UV Luminosity Function} \label{sec:Predict_UVLF}

Here, we describe our method for predicting the UVLF associated with potential LRD host galaxies and its comparison to observations under the different sets of modifications described in Section \ref{sec:mods}.  Note that the comparison performed here rests on the assumption that LRD UV emission is dominated by the host galaxies' stellar populations rather than by AGN.  This assumption is consistent with recent observations (e.g., \citealp{Killi_2024,Naidu_2025}) and some theoretical models (e.g., \citealp{Inayoshi_2025b}) of LRDs.

We extract intrinsic UV luminosities for each galaxy in our sample that has a predicted AGN bolometric luminosity in the luminosity range of LRDs ($L_{\rm Bol} = 10^{42.5-45.5} \rm \ erg/s$).  To do so, we model the rest-frame UV emission spectra of each galaxy's stellar component using BPASS v2.2 with nebular emission (lines and continuum; \citealp{Stanway_2018}).  We specifically convert to UV magnitudes from measurements of $L_{1566}$ where $L_{1566} = \langle \lambda L_{\lambda} \rangle$ averaged over $\lambda=1556-1576$ \AA.  To account for the effects of dust, we attenuate the UV luminosities predicted from each galaxy's stellar population according to the $z=6$ UV attenuation to stellar mass relation from \citet{Donnan_2025}.  Since it is uncertain whether the UV dust attenuation of LRD host galaxies follows this relation, we additionally explore the dust-free scenario in Appendix \ref{appendix:Dust}.  We then convolve our measured UV magnitudes with the HMF as described in Section \ref{sec:Calc_L_func} (but now binning by UV magnitudes rather than AGN bolometric luminosities).

\subsubsection{Effects of Modifications on Predicted LRD AGN Bolometric and LRD UV Luminosity Functions} \label{sec:Match_AGNLF}

Here, we discuss the effects that each modification to our model described above has on our predicted AGN bolometric luminosity function and LRD UVLF.  Specifically, we aim to relieve our fiducial models' overpredictions of the faint end of the LRD bolometric luminosity function ($L_{\rm Bol} \sim 10^{43-44} \rm \ erg/s$) and the LRD UVLF across its full range of observed UV magnitudes ($M_{\rm UV} = [-17, -21]$).  The top panels of Figure \ref{fig:LowL_Tension} present the AGN bolometric luminosity function, while the bottom panels present the UVLF for galaxies hosting AGN in the luminosity range of LRDs ($L_{\rm Bol} \sim 10^{43-45} \rm \ erg/s$) predicted following each set of modifications described in Section \ref{sec:mods}. 

Eliminating the contribution of low-mass galaxies (and therefore low-mass black holes) suppresses the faint end of the bolometric AGN luminosity function and UVLF.  Capping AGN luminosities at $L_{\rm Edd}$ slightly suppresses the bright end of the AGN bolometric luminosity function but has little effect on the predicted UVLF.  Including only super-Eddington accreting AGN whose bolometric luminosities are capped at $L_{\rm Edd}$ suppresses the luminosity functions across our luminosity range.  

Including only sources that are both above a minimum mass and are super-Eddington accreting can strongly suppress the bright end of the UVLF and the faint end of AGN bolometric luminosity function.  The bright end of the UVLF is suppressed in this case since contributions to the bright UV luminosity bins come from the most massive galaxies in our sample ($M_\star \gtrsim 10^{10} \ \rm M_\odot$) and it is difficult for the more massive black holes (presumed to live in these galaxies) to achieve super-Eddington accretion.  The faint end of the bolometric luminosity function is suppressed in this case since requiring that our galaxies, and therefore our black holes, be of a minimum mass and requiring super-Eddington accretion with $L_{\rm Bol}=L_{\rm Edd}$ effectively imposes a minimum bolometric luminosity ($L_{\rm Bol,\ Min}=L_{\rm Edd}(M_{\rm BH,\ Min})$).  This combination of effects is promising for matching both the observed peak in the LRD bolometric luminosity function at $L_{\rm Bol}\sim10^{44} \ \rm erg/s$ and the sharp decline at the bright end of the observed LRD UVLF.  

\begin{figure*}[!t]
    \centering
    \includegraphics[width=\linewidth]{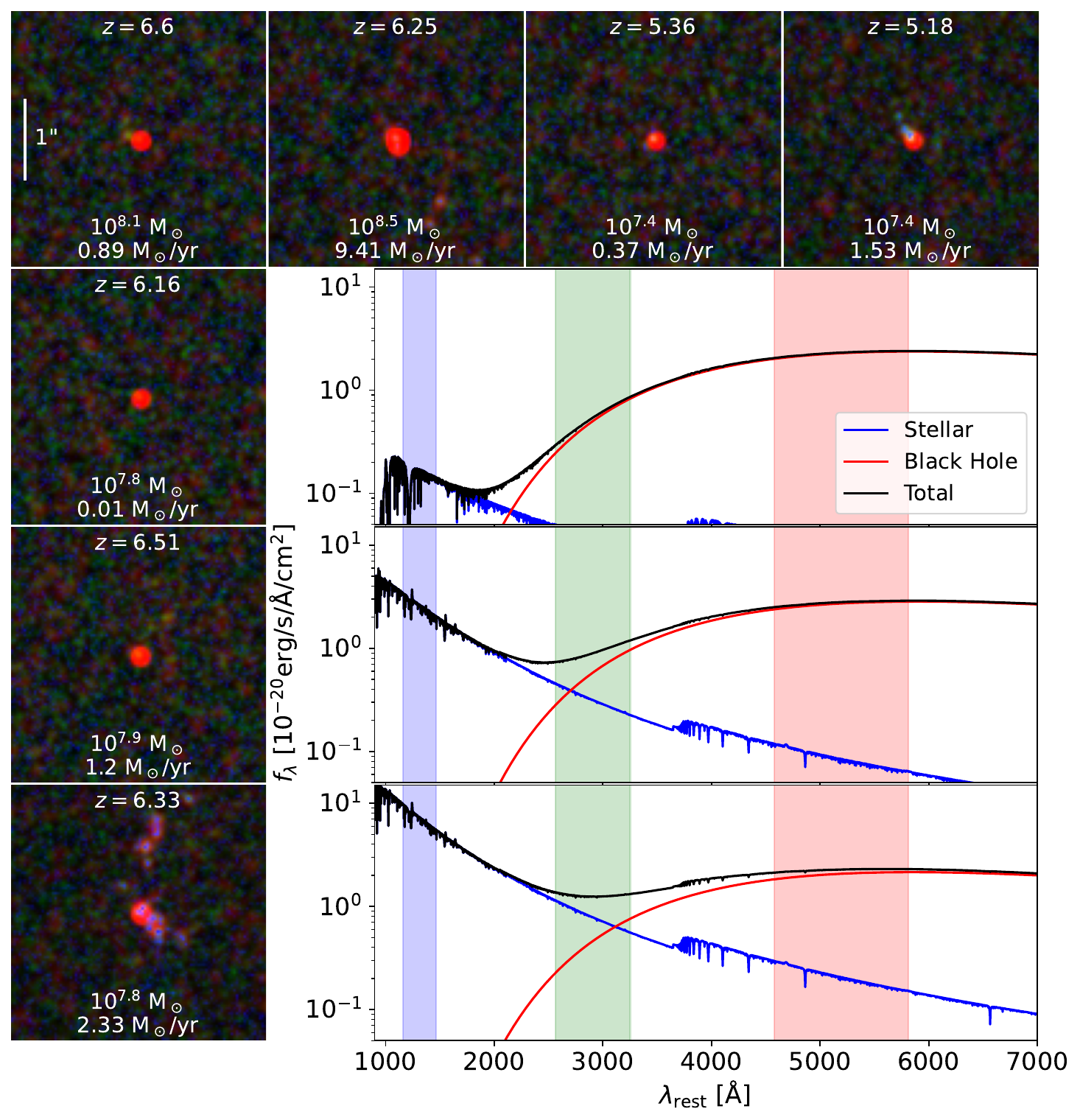}
    \caption{Mock $3''\times3''$ false-color JWST/NIRCam images of 7 plausible LRDs in the F090W, F200W, and F356W bands and 3 mock SEDs (to the right of their corresponding mock images).  Note $1''$ corresponds to a physical distance of $\sim6 \ \rm kpc$ at $z=6$  Stellar emission from within 1 kpc of each galaxy's center (blue) combined with an black hole-powered component (modeled here as a 5000 K blackbody; red) results in an overall ``v-shaped'' SED (black).  Image labels denote the galaxy's redshift (top), stellar mass (middle), and star formation rate averaged over the last 10 Myr (bottom).  The blue, green, and red highlighted bands in the spectra correspond to the wavelength coverage and respective colors of the F090W, F200W, and F356W bands in the mock RGB images.  Our plausible LRDs appear red and compact in the mock multi-band images.}
    \label{fig:mock_obs}
\end{figure*}

\subsubsection{A Plausible Scenario to Match LRD Demographics} \label{sec:Match_UVLF}

Focusing on the GTDA model, the set of modifications where we include only super Eddington-accreting, Eddington luminosity-limited AGN hosted within $M_\star>2\times10^{7} \ \rm M_{\odot}$ galaxies provides a promising simultaneous match to the shape of the observed LRD bolometric and LRD UV luminosity functions.  This scenario is denoted by the dotted green line in the left two panels of Figure \ref{fig:LowL_Tension}.  Note that, within the simple free-fall accretion model, this scenario predicts LRD bolometric and UV luminosity functions with shapes similar to those in the GTDA case; however, the GTDA model more closely reproduces the normalization of both observed relations.  We therefore refer to this scenario in the GTDA model, henceforth, as our ``Plausible LRD Scenario''. 

While the modifications made in this scenario are plausible and offer a promising simultaneous match to both the UV and AGN bolometric luminosity functions, we do not propose that they  provide hard constraints on the nature of LRDs.  There are too many model assumptions and other modifications one could make with potentially degenerate effects to draw conclusions on the physical origin of LRDs from this work alone.  Moreover, while predictions from our ``Plausible LRD Scenario'' qualitatively match the shape of the observed LRD bolometric and LRD UV luminosity functions, they do not quantitatively match observations in all bins.  This scenario does, however, illustrate the types of motivated cuts and physical hypotheses that one can impose on AGN and their host galaxies to bring our predictions into better agreement with LRD demographics.  It is also likely that there exists a fine-tuned set of modification parameters within the space explored here (i.e., more precise minimum stellar mass cut, maximum allowed Eddington luminosity ratio, and minimum Eddington accretion ratio cut) that would bring our predictions into closer quantitative agreement with observations.  For example, allowing for moderately super-Eddington luminosities and/or lowering the minimum Eddington accretion ratio may alleviate the underabundance of sources in the $L_{\rm Bol}\sim 10^{45}$ erg/s bin.

Note that our model that only excludes contributions from host galaxies with $M_\star<10^8 \ \rm M_\odot$ and makes no further modifications also provides a close match to the LRD bolometric luminosity function observed by \citet{Greene_2025}.  However, this model is unable to reproduce the shape of the LRD UVLF.  Moreover, this model is potentially in tension with clustering constraints presented by \citet{Matthee_2025} that suggest stellar masses of $M_\star \approx 5\times10^{7} \ \rm M_\odot$ for some LRDs and the recent observation of an LRD with an inferred host stellar mass of $M_\star \lesssim 10^{7.5} \ \rm M_\odot$ presented by \citet{Kokorev_2025}.  

\begin{figure*}[!t]
    \centering
    \includegraphics[width=\linewidth]{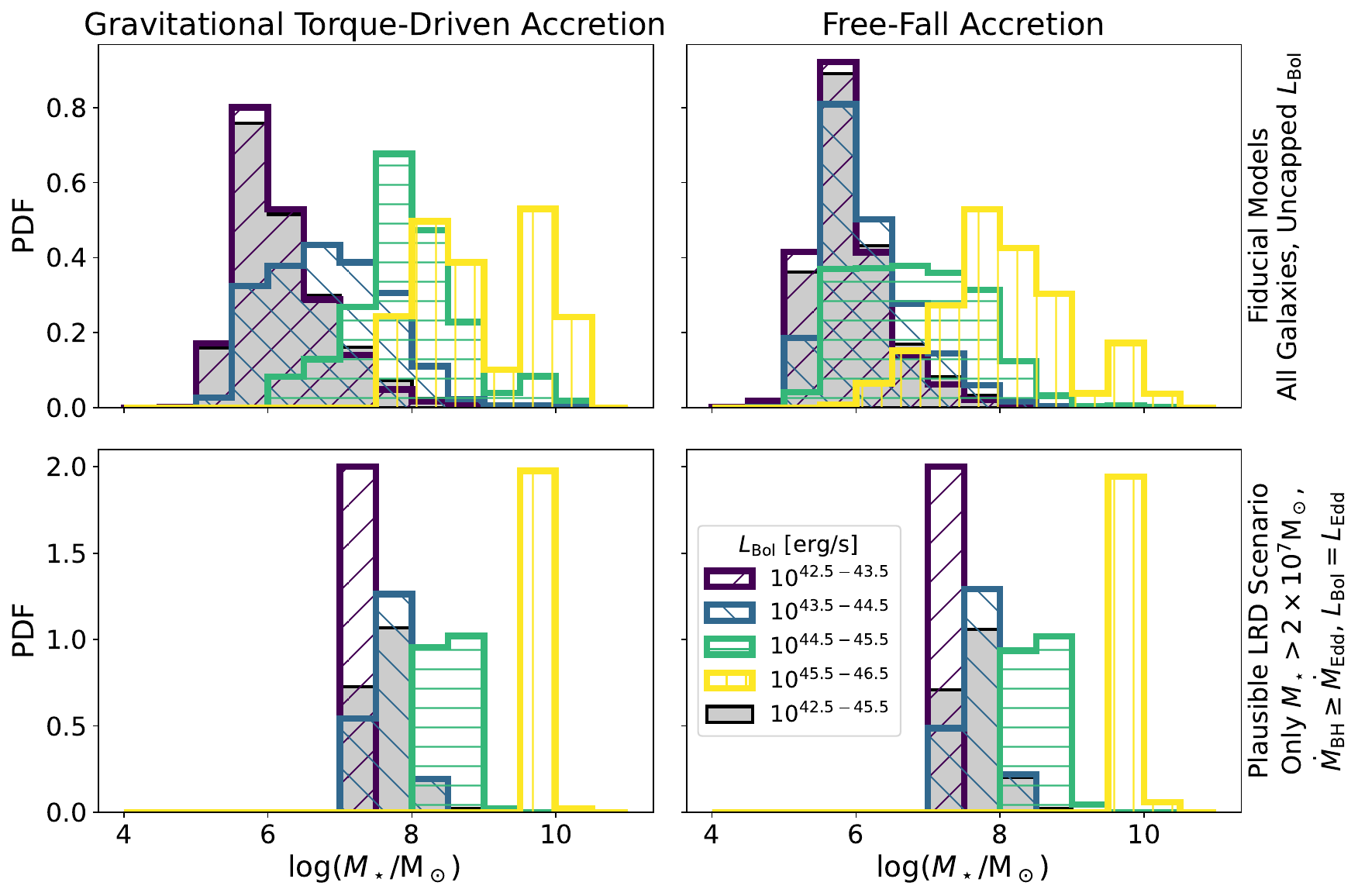}
    \caption{The stellar mass distribution of FIRE-2 galaxies hosting AGN in luminosity bins of $L_{\rm Bol} \sim 10^{42.5-43.5} \ (\rm purple), \ 10^{43.5-44.5} \ (\rm blue), \ 10^{44.5-45.5} \ (\rm green),  \ 10^{45.5-46.5} \ (\rm yellow), \ and \ 10^{42.5-45.5}$ (the full LRD luminosity range; black) erg/s} for our fiducial models (including unmodified $L_{\rm Bol}$ values for AGN in all galaxies; top) and our ``Plausible LRD Scenario'' (as in Figure \ref{fig:LowL_Tension}, including only super Eddington-accreting, Eddington luminosity-limited black holes hosted within $M_\star>2\times10^{7} \ \rm M_{\odot}$ galaxies; bottom) for the GTDA model (left) and the free-fall accretion model (right) at $z \sim 6$.  While more luminous AGN are typically hosted in more massive galaxies, our fiducial models predict AGN in each luminosity bin to have a wide range ($3-4$ orders of magnitude) in host stellar mass.  Our ``Plausible LRD Scenario'', however, predicts a much narrower range in host stellar masses of LRDs ($\sim 1$ order of magnitude) in a given luminosity bin and ($\sim 1.5$ orders of magnitude) when integrated over the entire LRD population.  In either case, the majority of AGN in the luminosity range of observed LRDs are hosted by low-mass ($\log(M_\star/\rm M_\odot) \lesssim 9$) galaxies whose central black holes accrete intermittently (see Figure \ref{fig:Accretion_Time_Series} for examples).
    \label{fig:M_star_Hist}
\end{figure*}

\subsection{Mock Observations of Plausible LRDs} \label{sec:mock_obs}

While we have demonstrated the ability of our ``Plausible LRD Scenario'' to qualitatively match observational demographics of LRDs, it is important to consider whether plausible LRD host galaxies in our sample can also reproduce their key observed characteristics (i.e., red and compact appearances, ``v-shaped'' SEDs).  Here, we demonstrate that galaxies selected as LRD hosts in our ``Plausible LRD Scenario'' can reproduce these key observational signatures following simple assumptions for the black hole-powered contribution to LRD spectra.  Figure \ref{fig:mock_obs} presents mock multi-band JWST NIRCam images and mock spectra of example plausible LRDs identified in the simulations.  Following our ``Plausible LRD Scenario'' above, plausible LRDs are selected to have $M_\star>2\times10^{7} \ \rm M_{\odot}$ and $\log(\dot{M}_{\rm Torque}/\dot{M}_{\rm Edd})>-0.5$ such that, following the application of our fiducial subgrid variability ($\sigma_{\log\epsilon}=0.5$), the GTDA model predicts super-Eddington accretion onto the central black hole for a minimum of 16 percent of realizations.  The sources presented in Figure \ref{fig:mock_obs} are selected to represent the redshift range of our analysis ($z=5-7$) and the stellar mass ($\log(M_\star/\rm M_\odot) \sim 7.3-8.5$) and star formation rate (averaged over the last 10 Myr; $\rm SFR_{10 Myr}\sim0.01-10 \ \rm M_\odot/yr$) ranges for our full sample of plausible LRDs.

We model the black hole-powered contribution as a $5000 \ \rm K$ blackbody placed at the center of the galaxy with bolometric luminosity equal to the Eddington luminosity for a $M_{\rm BH} = 0.01 M_\star$ black hole.  Phenomenologically, the optical and near-IR continuum of LRDs are well-approximated by $\sim 4000-6000 \rm \ K$ blackbody emission (e.g., \citealp{Kido_2025,Umeda_2026}).  This aligns with physical interpretations that LRDs are black holes embedded in dense gas cocoons that emit quasi-thermally (see e.g., \citealp{Naidu_2025, deGraaff_2025b, Rusakov_2026}).  We model the stellar continuum by inputting the properties of our simulated star particles into BPASS v2.3 (stellar continuum only; \citealp{Byrne_2022}).  Though all simulated stars in the viewing field contribute to our mock images, only stars within $1 \ \rm kpc$ of the galaxy's center contribute to our mock spectra.  This aperture approximately corresponds to the $0.2$ arcsec width of a NIRSpec slit at $z=6$.  We apply dust attenuation to our mock spectra following the $z=6$ UV attenuation to stellar mass relation from \citet{Donnan_2025} and assuming a dust optical depth versus wavelength relation $\tau_\lambda \propto \lambda^{-1}$.  We additionally apply the IGM transmission model from \citet{Inoue_2014} to our mock spectra.

We generate images by combining mock observations in the F090W, F200W, and F356W NIRCam bands with pixel sizes of 0.031", 0.031" and 0.063", respectively.  We apply Gaussian PSFs with FWHMs of 1.065, 2.129, and 1.841 pixels to these respective bands (values taken from the STScI NIRCam documentation; \citealp{Perrin_2014, Rigby_2023}).  We also add $27.5 \rm \ mag \ arcsec^{-2}$ of Gaussian noise to each band (approximately one fifth of the surface brightness corresponding to the $5\sigma$ point-source limiting magnitude of a JADES-Deep-like survey).  We apply dust attenuation to each filter using the same prescription from \citet{Donnan_2025} used for our mock spectra.  We then resize and crop the mock F356W image to match the pixel size and extent of the mock images in the other two bands before combining all three bands into a single false-color RGB image.

The F090W (blue; $\lambda_{\rm rest}(z=6)\sim1300$ \AA) filter is chosen to probe the rest-frame UV emission from recently formed, massive stars.  The F356W (red; $\lambda_{\rm rest}(z=6)\sim5100$ \AA) filter probes into the rest-frame optical where the spectrum is dominated by the black hole-powered component.  The F200W (green; $\lambda_{\rm rest}(z=6)\sim2850$ \AA) filter probes in between the other two filters, often near the inflection point of the ``v-shaped'' spectra.  Our plausible LRDs appear compact and red in our mock multi-band images, in agreement with photometric samples.  A few of our example plausible LRDs display extended UV emission, signaling extended star formation in the host galaxy.  For most, however, the UV emission is either compact or not detected altogether in our mock imaging.  Interestingly, we measure the underlying stellar population of our plausible LRD host galaxies to be quite extended (relative to what is detectable) with typical half-stellar mass radii of $\sim1-2 \ \rm kpc$.  This demonstrates the ability of extended LRD host galaxies to masquerade as compact sources in the UV by hiding extended stellar emission beneath the detection threshold of JWST NIRCam imaging surveys.

Furthermore, we find ``v-shaped'' mock SEDs (blue rest-UV and red rest-optical slopes) consistent with spectroscopic observations of LRDs.  The blue rest-UV slopes are due to stellar emission near the center of the galaxy while the red rest-optical slopes arise from the black hole-powered contribution.  This spectral decomposition is broadly consistent with recent observations (e.g., \citealp{Killi_2024,Naidu_2025}) and some theoretical models (e.g., \citealp{Inayoshi_2025b}) of LRDs.

We have shown that, under simple assumptions, our plausible LRD host galaxies can broadly reproduce key characteristics of observed LRDs.  However, more detailed mock observational modeling is necessary for precise observational comparison.  To begin, the rest-frame optical and NIR continua of LRD spectra are found to deviate from the simple blackbody assumed here, especially at the blue end of the rest optical.  Thus, modified blackbody fits have been shown to more precisely match LRD SEDs (e.g., \citealp{deGraaff_2025b}).  Moreover, significant Balmer breaks, which are neglected here, are present in many LRD spectra (e.g., \citealp{deGraaff_2025b}).  The deviation from simple blackbodies and the presence of Balmer breaks likely play a significant role in setting the inflection point of ``v-shaped'' LRD SEDs.  Finally, emission lines, which are also neglected here, appear prominently in observed LRD spectra.  Future work that self-consistently incorporates these spectral features via detailed radiative transfer calculations would allow for more comprehensive observational comparison.

\subsection{LRD Host Galaxy Stellar Masses} \label{sec:hostgalaxies}

Here, we present the stellar mass distribution of galaxies from our sample that host AGN in different luminosity bins and discuss the intermittent nature of black hole accretion in these galaxies.  We do so for both our fiducial models (including unmodified $L_{\rm Bol}$ values for AGN in all galaxies) and our ``Plausible LRD Scenario'' (including only super Eddington-accreting, Eddington luminosity-limited AGN hosted within $M_\star>2\times10^{7} \ \rm M_{\odot}$ galaxies) described above.  As done for our bolometric luminosity function analysis, we generate luminosities using 1000 bootstrapped resamples of accretion rates generated from our fiducial GTDA ($\bar\epsilon_{\rm T} = 5$, $\sigma_{\log\varepsilon_{T}} = 0.5$, $R = 100 \ \rm pc$) and free-fall accretion ($\bar\varepsilon_{\rm ff} = 0.01$, $\sigma_{\log\varepsilon_{\rm ff}} = 0.5$, $R = 100 \ \rm pc$) models.  Stellar masses are calculated as the total mass of all star particles within $0.2  R_{\rm vir}$ of the galaxies' centers.  Galaxies hosting AGN within given bolometric luminosity bins are then binned by stellar mass.  To make our sample representative of a cosmological volume, the contribution of galaxies to these bins are weighted according to the ``HMF-weighting'' scheme described in Section \ref{sec:Calc_L_func}.  

Figure \ref{fig:M_star_Hist} presents the probability that an AGN in a given bolometric luminosity bin ($L_{\rm Bol} \sim 10^{43} , \ 10^{44} , \ 10^{45} , \ \rm or \ 10^{46}  \ \rm erg/s$) is hosted within a galaxy of a given stellar mass.  As expected, more luminous AGN are more likely to be hosted by higher-stellar mass galaxies.  For our fiducial model, the host galaxies of AGN in all luminosity bins have a relatively wide range of stellar masses.  Our ``Plausible LRD Scenario'' predicts a much narrower range of stellar masses for LRD hosts.  Notably, in both sets of models, the vast majority of AGN in the luminosity range of observed LRDs from \citet{Greene_2025} ($L_{\rm Bol} \sim 10^{43-45}\ \rm erg/s$) are hosted by low-mass ($\log(M_\star/\rm M_\odot) \lesssim 9$) galaxies. As shown by examples in Figure \ref{fig:Accretion_Time_Series}, the central black holes of these low-mass galaxies accrete intermittently due to the disruption of the central gas supply by bursty stellar feedback.  \citet{AA_2017b}, \citet{Catmabacak_2022}, and \citet{Byrne_2023} have previously shown that black holes in FIRE simulations only transition to a more steady accretion phase at later times when their host galaxy is more massive ($M_\star \gtrsim 10^{10} \rm M_\odot$) and less bursty.  Nonetheless, we have demonstrated that the frequency and intensity of the intermittent black hole accretion within our population of bursty, low-mass galaxies is able to sustain an abundant population of luminous AGN at high redshift.

\subsection{Demographic Support for Super-Eddington Accretion LRD Models}

Since the discovery of LRDs, a number of theoretical models that involve super-Eddington accretion onto a central black hole have been put forth to explain their puzzling SEDs.  \citet{Madau_2024}, \citet{Inayoshi_2025c}, and \citet{Zhang_2025} demonstrate that super-Eddington accretion could naturally explain the observed X-ray weakness of LRDs.  \citet{Inayoshi_2025c} additionally find that a super-Eddington accretion scenario results in weak ultraviolet and optical variability, consistent with most observations of LRD variability, due to photon trapping within super-Eddington accretion disks.  Moreover, GRMHD simulations from \citet{Zhang_2025} demonstrate that super-Eddington accretion can result in the strong Balmer breaks observed in LRDs.  \citet{Inayoshi_2025} show that super-Eddington accretion models predict the high equivalent widths of broad H$\alpha$ lines and the presence of OI lines observed in LRDs.  Finally, a semi-empirical model developed by \citet{Lupi_2024} prefers a super-Eddington accretion scenario in order to match the broad-line and continuum emission of high-redshift AGN observed by JWST.  These models are consistent with recent broad and narrow line fitting performed on LRD spectra by \citet{Rusakov_2026}, which favors a low black hole mass ($M_{\rm BH}\sim10^{5-7} \ \rm M_\odot$), high Eddington ratio scenario for LRDs.

Until now, however, it has remained unknown whether galaxies are capable of supplying gas to their central black holes at super-Eddington rates frequently enough to explain the abundance of LRDs in the scenario that they are super-Eddington accreting AGN.  Our results, specifically the agreement of our ``Plausible LRD Scenario'' with observed LRD demographics, demonstrate that this is the case for galaxies evolving in a realistic cosmological environment, with stellar masses that are generally consistent with LRD observations ($M_\star>2\times10^{7} \ \rm M_{\odot}$), and with AGN luminosities limited at $L_{\rm Edd}$.  Moreover the simulations used here to demonstrate this point simultaneously reproduce other key high-redshift observations (i.e., the $z=8-12$ UVLF; \citealp{Sun_2023b} and the $z=5-12$ MZR; \citealp{Marszewski_2024,Marszewski_2025}).

\subsection{Limitations of the Present Work} \label{sec:limitations}

Here we discuss the limitations of this work, their potential effects, and how they may be improved in future work.  The first limitation is that we apply subgrid prescriptions to determine black hole accretion rates and their variability rather than truly resolving black hole accretion and its variability.  Looking forward, hyper-refined simulations that resolve black hole accretion on much finer scales in realistic cosmological environments (e.g., \citealp{AA_2021, Hopkins_2024c, Hopkins_2024a, Hopkins_2024b}) are a promising tool for understanding the nature of AGN and the limitations of the accretion models applied in this work.

A second limitation is that the simulations we analyze do not include AGN feedback, which, if included, would likely suppress accretion rates onto the black holes and potentially star formation in the galaxy.  If AGN feedback is capable of significantly disrupting accretion flows in the LRD regime, then our inferred accretion rates (and hence our luminosity functions) could be viewed as upper limits predicted from the simulations.  Since, over our luminosity range of interest, we largely overpredict the abundance of AGN, the inclusion of AGN feedback may bring our models into closer overall agreement with observations.  On the other hand, LRD observations notably do not detect any outflow or jet signatures (i.e., X-rays, e.g., \citealp{Yue_2024, Ananna_2024} and radio jets, e.g., \citealp{Akins_2025, Perger_2025}), hinting that AGN feedback could be suppressed in these sources.  Ultimately, simulations that include self-consistent, live-tracked black holes and AGN feedback are needed to elucidate the effects of AGN feedback on our results.   

\section{Conclusions} \label{sec:conclusions}
We have applied the GTDA and simple free-fall accretion models in post-processing to predict accretion rates for black holes at the center of bursty high-redshift FIRE-2 galaxies.  We have constructed AGN bolometric luminosity functions from these predicted accretion rates and we have predicted host galaxy UVLFs.  We compare our results with the observed demographics of LRDs and classical AGN.  Here, we summarize the key conclusions of this work:
\begin{itemize}
  \item Despite the highly intermittent black hole accretion rates inferred from our FIRE-2 sample, bursty, high-redshift galaxies are predicted to host a sufficiently abundant population of AGN in the luminosity range of LRDs.
  \item Assuming larger unresolved time variability in our predicted accretion rates has a net result of upscattering more galaxies from lower to higher luminosity bins.
  \item Our models predict that the vast majority of AGN in the luminosity range of observed LRDs are hosted by low-mass ($\log(M_\star/\rm M_\odot) \lesssim 9$) galaxies.
  \item Interestingly, our fiducial models reproduce the observed number of AGN at the highest LRD luminosities ($L_{\rm Bol} \sim 10^{45} \rm \ erg/s$) but (similar to other theoretical models in the literature) overpredict their abundance at lower luminosities ($L_{\rm Bol} \sim 10^{43-44} \rm \ erg/s$) relative to observations.  As a result, our fiducial models also overpredict the UVLF for host galaxies of AGN in the bolometric luminosity range of LRDs as compared with LRD observations.
  \item We explore potential explanations for these overpredictions and their implications for the types of galaxies that host LRDs and for black hole growth in the early universe.  A plausible (but likely not unique) scenario for simultaneously matching the LRD AGN bolometric luminosity function and the LRD UVLF suggests that LRDs are super Eddington-accreting, Eddington luminosity-limited, $M_{\rm BH}\gtrsim 2\times10^5 \ \rm M_\odot$ black holes residing in $M_\star \gtrsim 2\times10^7 \ \rm M_\odot$ galaxies.
  \item We generate mock observations (both multi-band images and SEDs) of galaxies selected as LRD hosts under this plausible scenario, assuming a simple 5000 K blackbody form for the black hole-powered component.  We demonstrate that these galaxies reproduce key observational signatures of LRDs (i.e., red and compact appearances, “v-shaped” SEDs) under this simple assumption.
\end{itemize}

This work is largely a ``first-order'' demographics check demonstrating the ability of bursty galaxies to host AGN in the luminosity range of LRDs in sufficient abundance.  More detailed theoretical work, as outlined in Section \ref{sec:limitations}, will help elucidate the true nature of LRDs.

Observationally, there are many outstanding questions that must be answered in order to achieve stronger demographic constraints on LRDs.  These key challenges include the separation of AGN and host galaxy contributions to their SEDs and more direct measurements of bolometric correction factors for a large sample of LRDs via observations over wide wavelength ranges (as done by \citet{Greene_2025} for two sources).

\bibliography{bib}{}
\bibliographystyle{aasjournal}

\section*{Acknowledgments}

AM was supported by a CIERA Board of Visitors Fellowship.  CAFG was supported by NSF through grants AST-2108230 and AST-2307327; by NASA through grants 80NSSC22k0809, 80NSSC22K1124 and 80NSSC24K1224; by STScI through grant JWST-AR-03252.001-A; and by BSF through grant no 2024262. GS was supported by a CIERA Postdoctoral Fellowship.  DAA acknowledges support from NSF CAREER award AST-2442788, NASA grant ATP23-0156, STScI grants JWST-GO-01712.009-A, JWST-AR-04357.001-A, and JWST-AR-05366.005-A, an Alfred P. Sloan Research Fellowship, and Cottrell Scholar Award CS-CSA-2023-028 by the Research Corporation for Science Advancement.  RF acknowledges financial support from the Swiss National Science Foundation (grant nos PP00P2-194814 and 200021-188552).  TBM was supported by a CIERA Fellowship.  The simulations analyzed in this work were run on XSEDE computational resources (allocations TG-AST120025, TG-AST130039, TG-AST140023, and TG-AST140064).  Analysis was done using the Quest computing cluster at Northwestern University.

\begin{appendices}

\section{Dependence on Radial Aperture of Black Hole Accretion Model} \label{appendix:radial_dependence}
\begin{figure*}[!t]
    \centering
    \includegraphics[width=\linewidth]{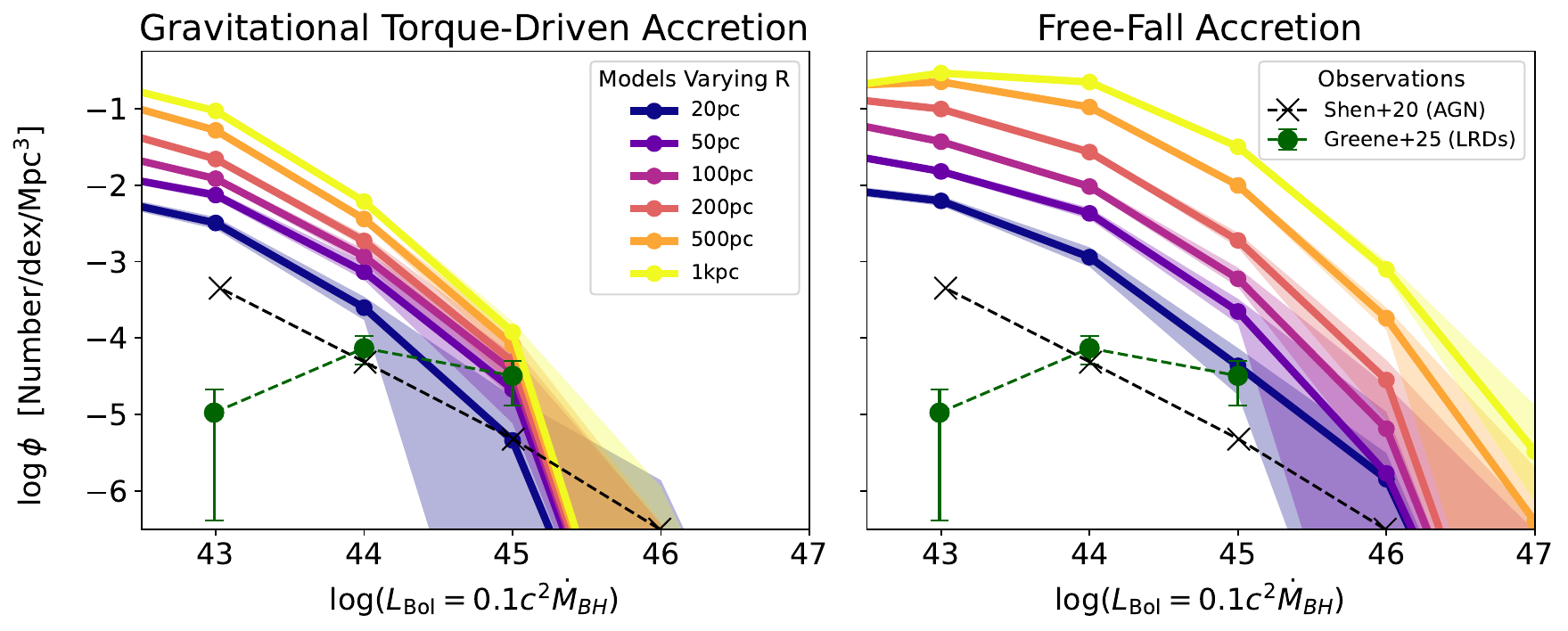}
    \caption{The predicted $z \sim 6$ AGN bolometric luminosity function using the GTDA model (left panel) and the simple free-fall accretion model (right panel) calculated using different radial apertures of $R = 20$ (dark blue), $50$ (purple), $100$ (magenta), $200$ (red), $500$ (orange), and $1000$ (yellow) pc.  Shaded regions represent the range between the 16th and 84th percentile luminosity functions predicted from our bootstrapped samples.  Green circles with error bars show the LRD bolometric luminosity function inferred from observations by \citet{Greene_2025} for $z=4-6$ and black crosses show a pre-JWST AGN bolometric luminosity function based on X-ray and UV observations from \citet{Shen_2020} (their ``Global Fit A'') for $z=4-6$.  Here, we introduce our fiducial amount of variance to the normalization/accretion efficiency ($\sigma_{\log\epsilon_T} = \sigma_{\log\varepsilon_{\rm ff}}=0.5$) and we use our fiducial accretion efficiency of $\varepsilon_{\rm ff}=0.01$ for the simple free-fall model.  The predicted luminosity function using the free-fall accretion model increases when using larger radial apertures.  The luminosity function predicted from the GTDA model is less sensitive to the choice in aperture, especially at the bright end.  We choose $R = 100$ pc as our fiducial radial aperture for both models since this aperture remains significantly larger than the typical resolution (median smoothing length) achieved for the gas in the central regions of our simulated galaxies ($\sim 10-20$ pc).}
    \label{fig:AGN_Bol_LF_vs_R}
\end{figure*}

Here, we explore the dependence of our predicted AGN bolometric luminosity functions on $R$, the radial aperture within which the properties of particles are used to predict the accretion rate of central black holes.  We repeat our procedure to calculate the luminosity function using both the GTDA and free-fall accretion models using the properties of particles located within different radial apertures of $R = 20$, $50$, $100$, $200$, $500$, and $1000$ pc.  In Figure \ref{fig:AGN_Bol_LF_vs_R} we present the dependence of the AGN luminosity function predicted from the GTDA and free-fall accretion models on $R$.  We find that the bolometric luminosity function predicted from the simple free-fall accretion model increases significantly with larger $R$ due to the increased gas mass available within larger apertures.  However, our results from the GTDA model are relatively insensitive to aperture size (especially at the bright end, where a priori the LRD population may be most challenging to explain in bursty galaxies), with larger apertures resulting in only slightly higher predicted luminosity functions.  The relative insensitivity of the GTDA model on aperture size is a desirable property and another reason our exploration of plausible physical scenarios in Section \ref{sec:Match_Obs} focuses on this model.

Since the process of accretion onto a black hole takes place on scales much smaller than are resolved by our simulations, it is desirable to calculate predicted accretion rates from properties of particles within the smallest possible aperture that still typically captures the dynamics of many gas particles.  In the lowest resolution simulations used in this work, we find the median smoothing lengths of gas particles in the central region of galaxies to be $\sim 10-20$ pc. We therefore choose $R=100$ pc as our fiducial aperture.

\section{Effects and Self-Consistency of Assumed $M_{\rm BH}-M_\star$ Relation} \label{appendix:MBH_Assumption}

Throughout this work we have assumed a $M_{\rm BH}-M_\star$ scaling relation of $M_{\rm BH} = 0.01M_\star$.  This fiducial relation is in approximate agreement with the mean relation presented by \citet{Harikane_2023} for $z=4-7$ galaxies.  However, there is a significant scatter about this observationally derived relation and the true relation remains difficult to constrain due to observational uncertainties and systematics.  Our results obtained from the fiducial GTDA model are insensitive to this assumed scaling relation due to the inherently weak scaling of the accretion rates in this model with black hole mass ($\dot{M}_{\rm Torque}\propto M_{\rm BH}^{1/6}$).  Likewise, accretion rates predicted from the fiducial simple free-fall model are insensitive to this choice since the black hole mass typically makes up only a small fraction of the total mass within the aperture ($M_{\rm BH} \ll M_{\rm tot}$, except for in extreme cases).  However, our modified models that invoke the Eddington (luminosity and/or accretion) limits are sensitive to this choice due to the linear scaling of $L_{\rm Edd}$ and $\dot{M}_{\rm Edd}$ with $M_{\rm BH}$.  In this appendix, we therefore investigate the effects of assuming different $M_{\rm BH}-M_\star$ scaling relations ($M_{\rm BH} = 0.1M_\star$ and $M_{\rm BH} = 0.001M_\star$) on the results derived from these modified models.  We also test the self-consistency of our fiducial relation by checking whether integrating accretion rates from the GTDA and free-fall models reproduces black hole growth along this scaling relation.

\begin{figure*}[!t]
    \centering
    \includegraphics[width=\linewidth]{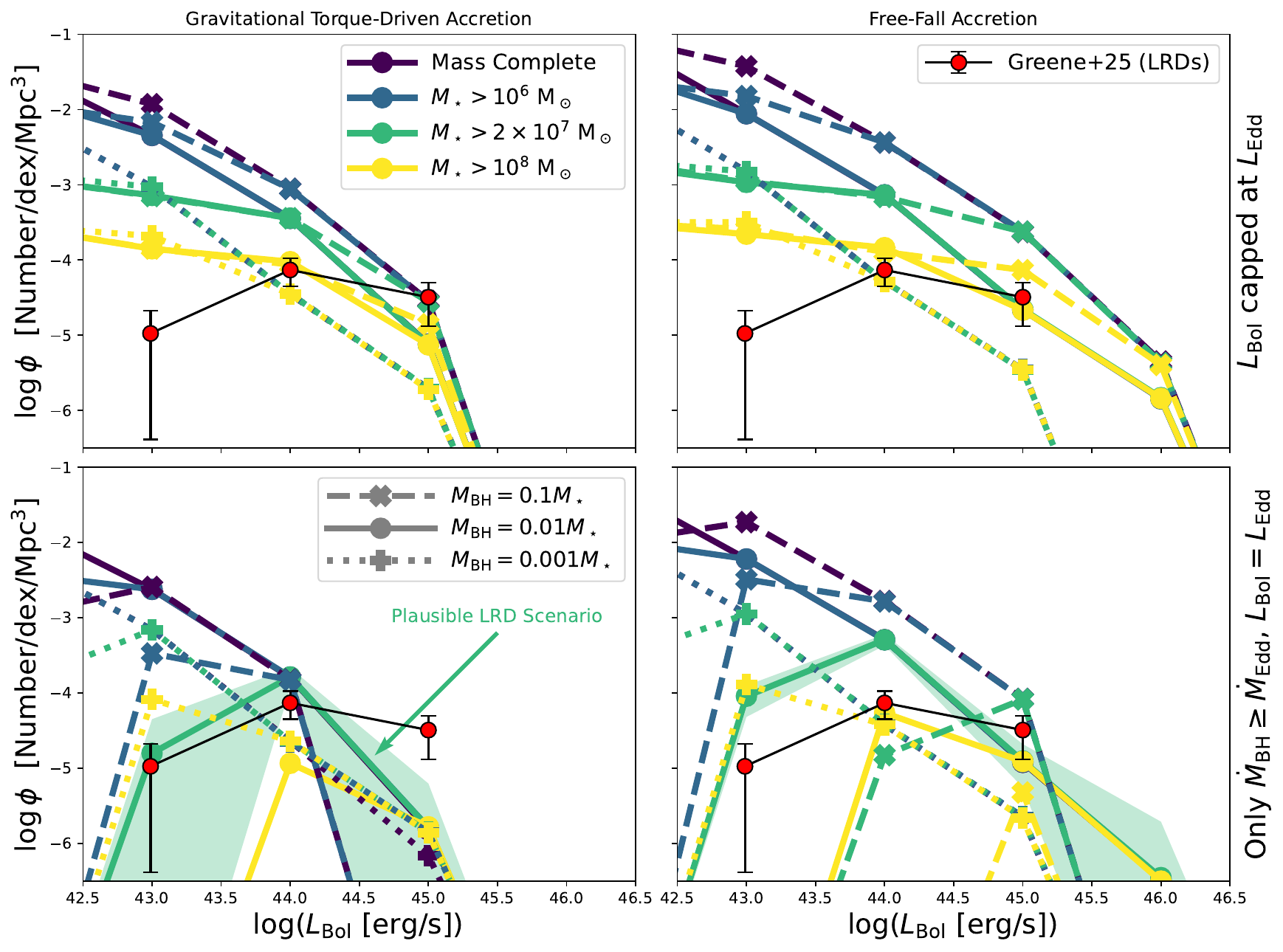}
    \caption{The predicted $z \sim 6$ AGN bolometric luminosity function assuming different $M_{\rm BH}-M_\star$ scaling relations for the cases where the luminosities of AGN are capped at their predicted $L_{\rm Edd}$ (top) and where we only include super-Eddington accretors whose luminosities are still capped at $L_{\rm Edd}$ (bottom) using the GTDA model (left) and the simple free-fall accretion model (right).  We present the cases where galaxies of any mass are permitted to host AGN (dark purple) and where the contributions of AGN hosted within galaxies with $M_\star<10^{6} \ \rm M_{\odot}$ (blue), $M_\star<2\times10^{7} \ \rm M_{\odot}$ (green), and $M_\star<10^{8} \ \rm M_{\odot}$ (yellow) are excluded.  In addition to our fiducially assumed scaling relation of $M_{\rm BH} = 0.01M_\star$ (solid lines), we also test $M_{\rm BH} = 0.1M_\star$ (dashed lines) and $M_{\rm BH} = 0.001M_\star$ (dotted lines).  Red circles show observations of the $z=4-6$ LRD bolometric luminosity function from \citet{Greene_2025}.  Our ``Plausible LRD Scenario'' is shown here by the solid green line with shaded region representing the range between the 16th and 84th percentile luminosity functions from our bootstrapped samples.  In the case where we only cap luminosities at $L_{\rm Edd}$, larger assumed black hole masses result in higher AGN bolometric luminosity functions, especially at the bright end.  However, in the case that we also only allow super-Eddington accreting black holes to contribute, larger assumed black hole masses can decrease our predicted luminosity function (especially at the faint end) due to difficulty in reaching the elevated Eddington rates.}
    \label{fig:BHLF_vs_MBH_assumption}
\end{figure*}

Figure \ref{fig:BHLF_vs_MBH_assumption} presents the $z \sim 6$ AGN bolometric luminosity function  assuming different $M_{\rm BH}-M_\star$ scaling relations for our modified models that invoke Eddington luminosities and/or accretion rates.  In the case where we only cap luminosities at $L_{\rm Edd}$, larger assumed black hole masses result in a slightly higher AGN bolometric luminosity function, especially at the bright end.  However, in the case that we also only allow super-Eddington accreting black holes to contribute to the LRD population, larger assumed black hole masses can decrease our predicted luminosity function since very few black holes are predicted to accrete at the elevated Eddington rates.  Therefore, if $M_{\rm BH}/M_\star \gtrsim 0.01$, then the bright end of the LRD bolometric luminosity function becomes difficult to explain in super-Eddington scenarios.

\begin{figure*}[!t]
    \centering
    \includegraphics[width=\linewidth]{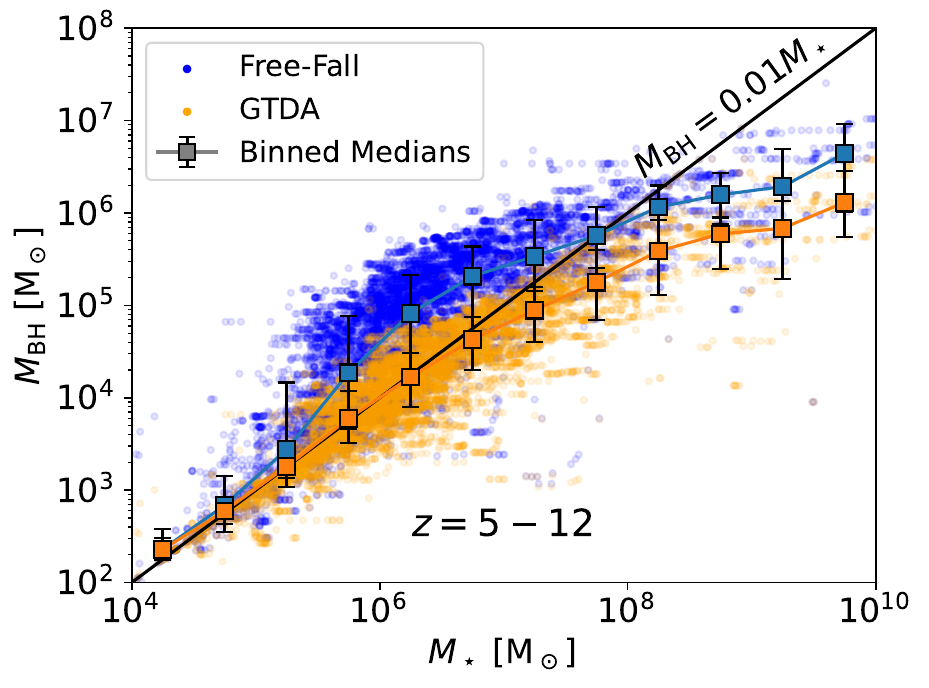}
    \caption{Predicted $M_{\rm BH} - M_\star$ scaling relations for all galaxies tracked in our analysis for $z=5-12$ in a simplified growth scenario.  Black hole masses are calculated by integrating the growth predicted from the GTDA (orange) and free-fall accretion (blue) models.  Black holes are given initial seed masses according to our fiducially assumed scaling relation ($M_{\rm BH} = 0.01M_\star$) at the first timestep for which they are tracked.  Boxes show stellar mass-binned median values with error bars representing the 16th and 84th percentiles.  Black holes grown via the GTDA model remain on our fiducially assumed scaling relation until their host galaxies reach $M_\star \sim 10^7 \ \rm M_\odot$, after which they gradually become undermassive.  Typical black holes grown via the free-fall accretion model are moderately overmassive until their host galaxies grow to be $M_\star \sim 10^{8} \ \rm M_\odot$, after which they gradually become undermassive.  Black holes in the galaxy mass range of our plausible LRDs ($M_\star \sim 10^{7.3-8.5} \ \rm M_\odot$) do not deviate greatly from our assumed relation despite our simple model neglecting black hole growth through mergers.}
    \label{fig:BH_Growth}
\end{figure*}

Figure \ref{fig:BH_Growth} presents the $M_{\rm BH}-M_\star$ scaling relations predicted from the GTDA and free-fall accretion models.  Initial black hole masses are seeded along our fiducially assumed scaling relation ($M_{\rm BH} = 0.01M_\star$) at the first snapshot for which each galaxy has $M_\star>10^4 \ \rm M_\odot$ over the interval $z=5-12$.  For each galaxy at each subsequent snapshot, the black hole is grown by integrating the accretion rates predicted from our GTDA and free-fall accretion models.  We find that typical black holes grown via the GTDA model remain on our fiducially assumed scaling relation until their host galaxies reach $M_\star \sim 10^7 \ \rm M_\odot$, after which they gradually become undermassive relative to this relation.  Typical black holes grown via the free-fall accretion model are predicted to be moderately overmassive for host galaxies with $M_\star \sim 10^{5.5-7.5} \ \rm M_\odot$, fall on the scaling relation for $M_\star \sim 10^8 \ \rm M_\odot$ galaxies, and gradually become undermassive for galaxies above this.  Note that these results should not be taken as detailed predictions as we have made simple assumptions regarding black hole seed mass and location and have neglected the growth of black holes through mergers (which, if included, may reduce the extent to which black holes are undermassive in more massive galaxies).  Even so, black hole masses do not deviate greatly from our fiducial scaling relation for galaxies in the stellar mass range of our plausible LRDs ($M_\star \sim 10^{7.3-8.5} \ \rm M_\odot$; see Figure \ref{fig:M_star_Hist}).  This suggests that our ``Plausible LRD'' scenario could be self-consistently realized.

\section{Dust Effects on LRD UVLF} \label{appendix:Dust}

\begin{figure*}[!t]
    \centering
    \includegraphics[width=\linewidth]{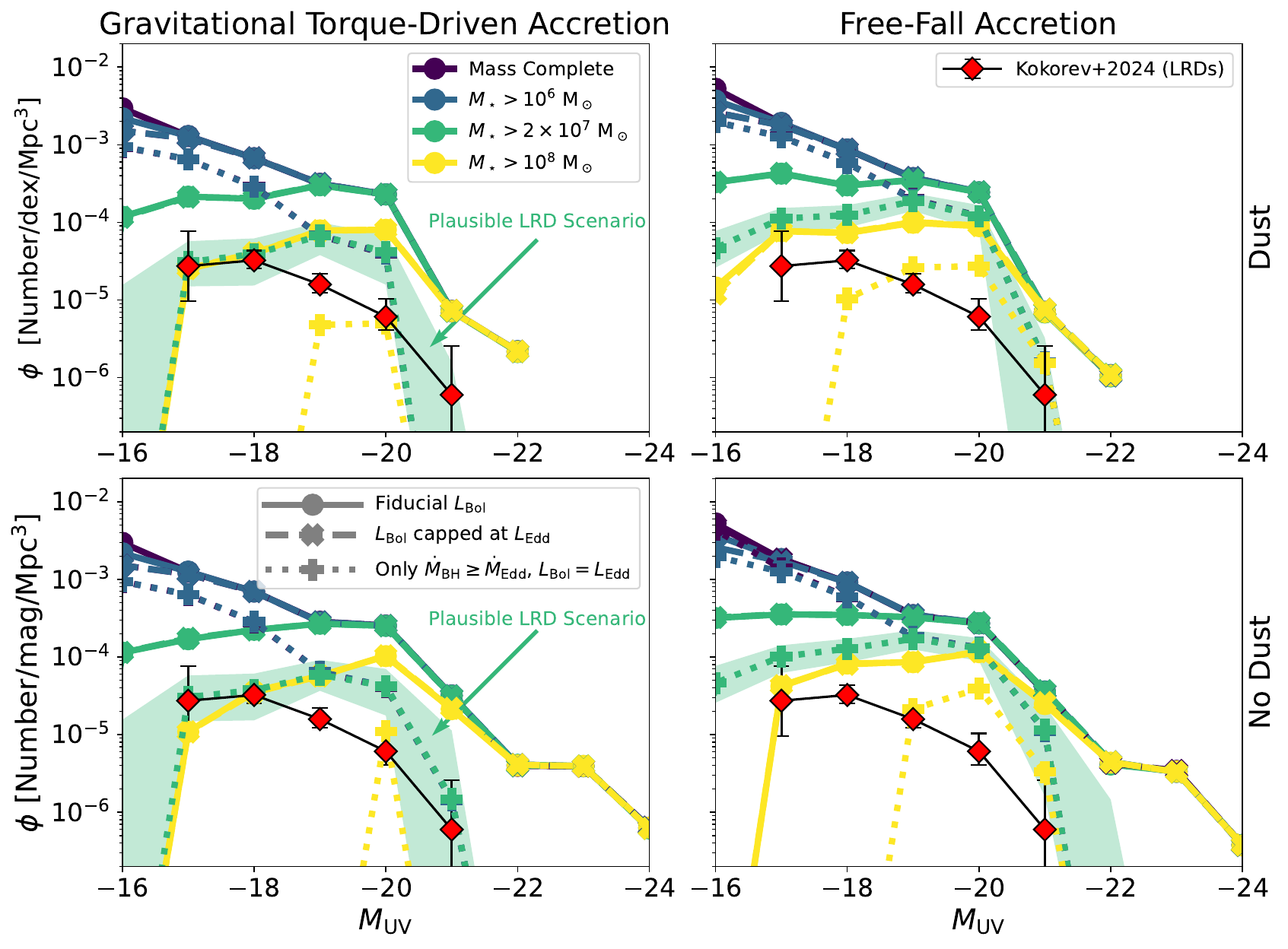}
    \caption{The predicted $z \sim 6$ UVLF for galaxies hosting an AGN in the LRD luminosity bins ($L_{\rm Bol} \sim 10^{43-45} \ \rm erg/s$) using the GTDA model (left) and the simple free-fall accretion model (right) using our fiducial dust model from \citet{Donnan_2025} (top) and a dust-free model (bottom).  Galaxies only contribute to the UVLFs if they are predicted (by the model plotted) to host an AGN in the range $L_{\rm Bol} = 10^{42.5-45.5} \ \rm erg/s$.  In these models, the UV luminosity is powered by star formation in the host galaxy only.  We present the cases where galaxies of any mass are permitted to host AGN (dark purple) and where the contributions of AGN hosted within galaxies with $M_\star<10^{6} \ \rm M_{\odot}$ (blue), $M_\star<2\times10^{7} \ \rm M_{\odot}$ (green), and $M_\star<10^{8} \ \rm M_{\odot}$ (yellow) are excluded.  With these host galaxy stellar mass cuts, we simultaneously explore the cases where the luminosities of AGN are capped at their predicted $L_{\rm Edd}$ (dashed) and where we only include super-Eddington accretors whose luminosities are still capped at $L_{\rm Edd}$ (dotted).  Our ``Plausible LRD Scenario'', in which LRDs are powered by black holes accreting at super-Eddington rates, is shown by the dotted green line with the shaded region representing the range between the 16th and 84th percentile luminosity functions from our bootstrapped samples.  Red diamonds show the $z=4.5-6.5$ LRD UVLF from \citet{Kokorev_2024}.  For all models, the primary effect of including dust is its suppression of the bright end of the UVLF ($M_{\rm UV}\lesssim-21$).  Dust attenuation also has a significant effect on our set of models that only include $M_\star>10^{8} \ \rm M_{\odot}$ galaxies as potential LRD hosts (yellow curves) across our full luminosity range.  Both of these effects are consistent with the form of the UV attenuation to stellar mass relation we employ from \citet{Donnan_2025}, which predicts more massive galaxies to have more dust attenuation.}
    \label{fig:BHLF_vs_Dust}
\end{figure*}

In Section \ref{sec:Predict_UVLF} and \ref{sec:Match_AGNLF} we present our modelling and predictions for the LRD UVLF.  This modelling assumed that the UV luminosities of LRD host galaxies are dust-attenuated according to the $z=6$ UV attenuation to stellar mass relation from \citet{Donnan_2025}.  Here we present the predicted LRD UVLF under an alternative, dust-free scenario and discuss the effects of including dust attenuation.  While the effect of dust on the spectra of LRD hosts is still uncertain, the dust-free scenario is motivated by the lack of observed dust-reprocessed emission in the far-infrared spectra of LRDs (e.g., \citealp{Setton_2025,Casey_2025}).  Figure \ref{fig:BHLF_vs_Dust} presents the predicted $z \sim 6$ LRD UVLF for all models described in Section \ref{sec:mods} with and without dust attenuation.

The primary effect of including-dust attenuation is its suppression of the bright end of the UVLF ($M_{\rm UV}\lesssim-21$).  This is a result of the most massive galaxies in our sample, which are the dominant contributors to the bright end of the UVLF, being the most severely dust-attenuated according to the \citet{Donnan_2025} relation.  For the same reason, the UVLFs for our set of models that only include $M_\star>10^{8} \ \rm M_\odot$ galaxies as potential LRD hosts are suppressed by dust attenuation across the full luminosity range.  For all other models explored here, dust attenuation does not have a significant impact on the faint end of the UVLF ($M_{\rm UV}\gtrsim-21$).

\end{appendices}

\end{document}